\documentclass[conf]{new-aiaa} %for journal papers
\usepackage{graphicx}
\usepackage{amsmath}
\usepackage[version=4]{mhchem}
\usepackage{siunitx}
\usepackage{subfigure}
\usepackage{longtable,tabularx}
\usepackage{lineno}

 \title{Reynolds number dependence of length scales governing turbulent flow separation with application to wall-modeled large-eddy simulations}
\author{Rahul Agrawal \footnote{Ph.D. candidate, Department of Mechanical Engineering. Student Member AIAA. (Email: rahul29@stanford.edu) }}
\affil{Center for Turbulence Research, Stanford University}
\author{Sanjeeb T. Bose \footnote{Distinguished Engineer, Cadence Design Systems and Adjunct Professor, Institute for Computational and Mathematical Engineering, Stanford University. Senior Member AIAA. }}
 \affil{Cadence Design Systems, Inc.}
\affil{Institute for Computational and Mathematical Engineering, Stanford University }
\affil{Center for Turbulence Research, Stanford University}
\author{Parviz Moin \footnote{Franklin P. and Caroline M. Johnson Professor, Dept. of Mechanical Engineering, Stanford University. Director of the Center for Turbulence Research. Fellow AIAA. }}
\affil{Center for Turbulence Research, Stanford University}

\date{\today}
\begin{document}

\maketitle

\begin{abstract}
This work proposes a Reynolds number scaling of the required number of grid points to perform wall-modeled LES of turbulent flows encountering separation off a solid surface. Based on comparisons between the various time scales in a non-equilibrium turbulent boundary layer (due to the action of an external pressure gradient), a simple definition of the near-wall ``under-equilibrium" and ``out-of-equilibrium" scales is put forward (where ``under-equilibrium" scales are governed by a quasi-balance between the viscous and the pressure gradient terms).  It is shown that this ``under-equilibrium" characteristic length scale varies with Reynolds number as $l_p \sim Re^{-2/3} $. The same scaling is obtained from a simplified Green's function solution of the Poisson equation in the vicinity of the separation point. \textit{A-priori} analysis is used to demonstrate that the resolution required to reasonably predict the wall-shear stress (for example, errors lower than approximately $10-15\%$ in the entire domain) in several nonequilibrium flows is at least $ \mathcal{O}(10) $$l_p$ irrespective of the Reynolds number and the Clauser parameter. Further, a series of \emph{a-posteriori} validation studies are performed to determine the accuracy of this scaling including the flow over the Boeing speed bump, Song-Eaton diffuser, Notre-Dame Ramp, and the backward-facing step. The results suggest that for these flows, scaling the computational grids ($\Delta$) such that $\Delta/l_p$ is independent of the Reynolds number results in accurate predictions of flow separation at the same ``nominal" grid resolution across different Reynolds numbers. Finally, it is suggested that atleast locally, in the vicinity of the separation and reattachment points, the grid-point requirements for wall-modeled large eddy simulations may scale as $Re^{4/3}$, which is more restrictive than the previously proposed flat-plate boundary layer-based estimates ($ \sim Re^{1}$) of Choi and Moin, \emph{Phys. Fluids, 2012} and Yang and Griffin, \emph{Phys. Fluids, 2021}.  
\end{abstract}

\section{Nomenclature}

{\renewcommand\arraystretch{1.0}
\noindent\begin{longtable*}{@{}l @{\quad=\quad} l@{}}
$\alpha$ & angle of tapering \\
$\rho$ & density \\
$e$ & resolved internal energy \\
$E$ & sum of resolved internal and kinetic energies \\
$T$ & temperature \\
$p$ & pressure \\
$u_i$ & velocity components \\
$S^d_{ij}$ & deviatoric component of the strain-rate tensor \\
$\tau^{sgs}_{ij}$ & subgrid-scale stress tensor \\
$Q^{sgs}_{j}$ & subgrid heat-flux \\
$l_p$ & pressure-gradient imposed viscous length scale \\
$y^+$ & resolution based on friction velocity \\
$U_{\infty}$ & freestream velocity \\
$\delta$ & boundary layer thickness \\
$\theta$ & momentum thickness \\
$\Delta$ & size of computational grid \\
$u_{\tau}$ & friction velocity \\
$u_p$ & pressure-gradient based velocity \\
$Pr$  & Molecular Prandtl number  \\
$Pr_{t}$ & Turbulent Prandtl number  \\
$Re_L, \; Re_{H}$ & Reference Reynolds number based on a geometric length scales \\
$Re_{\tau} $ & Friction Reynolds number \\
$Re_{\theta}^{ref}$ & Reference Reynolds number based on momentum thickness \\
$t_p$ &  time-scale based on balance pressure-gradient and viscous terms \\
$ t_v$ &  time-scale based on near-wall viscous effects  \\
$t_{iv}$ &  time-scale based on local-shear   \\
$Mcv$ & Million Control Volumes \\
$LES$ & Large-eddy simulation \\

\end{longtable*}}

\section{Introduction}
%\vspace{-5pt}
In complex engineering flows, the boundary layers over solid surfaces often experience a combination of favorable and adverse pressure gradients. A favorable pressure gradient reduces the turbulent normal stresses in the near-wall flow, making it more organized \citep{sreenivasan1982laminarescent}. In the presence of an adverse pressure gradient, however, these normal stresses near the wall become stronger \citep{adams1988flow}. In the latter case,  these non-equilibrium flows can even exhibit a boundary layer separation from the surface before finally reattaching downstream. The structure of turbulent separation bubbles has remained a matter of active research \citep{lighthill1953boundary,kiya1983structure,adams1988flow,wu2005low,hudy2007stochastic,mohammed2016unsteadiness,wu2020total}. Even in canonical problems of boundary layers experiencing pressure gradients without separating off the surface, integrated streamwise history effects have been reported \citep{bobke2017history} to determine the downstream state of a boundary layer. \citet{devenport2022equilibrium} have suggested that the flow history effects may even persist up to 50 local boundary-layer thicknesses downstream. Recently, \citet{agrawal2023thwaites} developed an extension of the method of Thwaites which also serves as a quantitative analysis tool to account for the flow history.

%%---------------------------------------
% stb: i am not sure that this paragraph adds much.  it is aircraft 
% specific and its relation to the introductory paragraph is tenuous.
% i'd recommend we remove it - but i leave it to you
%%--------------------------------------

%Existing results from wall-modeled LES for the flow over the Common Research Model aircraft \citep{agrawalarb2023a,kiris2022high,goc2023wind,kiris2023hlpw} have demonstrated that converging the predictions (of the pitching moment) from simulations at the higher angles of attack (near, at, and post-stall region) is more challenging than at the lower angles of attack where the flow is primarily attached. %This effect was also ascribed to the stronger flow acceleration at the leading edges of the aircraft at the higher angles of attack. An \textit{``a-posteriori"} Reynolds number scaling of the grid-point requirements for simulating such flows is yet to be reported. \citet{agrawalarb2023a} have recently presented significant grid-sensitivities in the solutions near stall as the Reynolds number increases.

Detailed knowledge of the near-wall structure of equilibrium boundary layers has been leveraged to establish resolution requirements for both wall-resolved and wall-modeled large-eddy simulations (LES). For example, a resolution requirement of $\Delta u_{\tau} /\nu \sim O(1)$ (where $\Delta$ is the size of the near-wall grid-cell, and $u_{\tau}$ is the friction-velocity) is recommended for wall-resolved LES or the validity of the law of the wall in the logarithmic regions of the velocity profiles in wall modeled LES.  Despite the augmentation of the structure of the turbulent boundary layer in the presence of strong pressure gradients, existing grid point estimates still rely on the phenomenology of equilibrium boundary layers \citep{chapman1979computational,choi2012grid,yang2021grid}.  
These studies suggest that the grid point requirements for wall-modeled LES scale linearly with the Reynolds number if resolution with respect to the boundary layer remains fixed.  There are two 
shortcomings of this approach.  First, the conclusions regarding the Reynolds number scaling may be incorrect in the presence of 
pressure gradients.  Second, and perhaps more importantly, the existing grid point estimates do not provide estimates on the \emph{a-posteriori} errors in quantities of interest given a specified outer layer resolution.  For instance, \citet{lozano2022performance} show that errors in the prediction of mean velocity profiles greatly increase in regions of adverse pressure gradients compared to zero pressure gradient regions even if the outer layer resolution is fixed.  This is especially true in separating boundary layers, where significantly higher outer layer resolution has been required for accurate simulations of flow exhibiting both smooth body separation \citep{whitmore2022brief} and large, three-dimensional separation patterns \citep{agrawalarb2023a,kiris2022high,goc2023wind,kiris2023hlpw}. It is likely that for such flows, the distribution of the pressure gradient may affect the convergence rate of wall-modeled LES toward reference quantities of interest. An example is the flow over an aircraft model, where for different angles of attack, the pressure gradient distribution changes due to a change in the effective body shape. Under milder pressure gradients (at low angles of attack), the flow exhibits smooth body separation over the trailing edge of the flaps, and for stronger pressure gradients (at high angles of attack), the flow can exhibit strong unsteadiness and large separation patterns emanating from the leading edges. It has been reported \citep{goc2023wind,kiris2023hlpw} that converging the lift, pitching moment and integrated drag for different angles of attacks requires different grid resolutions for the same nominal Reynolds number flow.

This article aims to address the two aforementioned shortcomings, specifically, in the context of separating flows. A characteristic length scale that governs the near-wall flow equilibrium (a quasi-balance between the viscous and the pressure gradient terms) is used to determine the resolution required for accurately predicting quantities of interest for flows (such as the surface pressure) that exhibit a smooth body separation across a range of Reynolds numbers. To test the validity of this scaling, we use the charLES flow solver \citep{bres2018large} to perform wall-modeled LES. The rest of the article is organized as follows. In section III we describe the governing equations being solved in charLES solver and the choice of the subgrid-scale and the wall model employed in this work. Section IV discusses the physical length scale that governs the onset of flow separation. Section V discusses \emph{a-posteriori} wall-modeled LES results across multiple flows. Finally, some conclusions are drawn in Section VI.

\section{Numerical Solver and Governing equations}
%\vspace{-5pt}

\noindent
The simulations presented in this work are performed using charLES, an explicit, unstructured, finite-volume solver for the compressible Navier-Stokes equations. This code is formally 2\textsuperscript{nd}-order accurate in space, and 3\textsuperscript{rd}-order accurate in time, and utilizes grids based on Voronoi diagrams. More details of the solver and validation cases can be found in \citet{agrawal2023wall,bres2018large,goc2021large,fu2021shock}. Formally skew-symmetric operators are employed to conserve kinetic energy, these discrete operators also approximately preserve entropy in the inviscid, adiabatic limit.  \\

A brief description of the previously mentioned compressible Navier-Stokes equations for LES is provided as follows. If the filtered, large-scale fields (such as velocity and pressure) in LES are denoted by $\overline{f}$, and their corresponding Favre average is denoted by $\widetilde{f}$, then the resulting equations of a compressible flow [of internal energy $e$, density $\rho$,  temperature $T$, viscosity $\mu(T)$ and thermal conductivity $\kappa(T)$] and velocity vector $\vec{u} \; = \; \{ u_1, \;u_2, \;u_3\}$ are given as, 

\begin{equation}
\frac{\partial \overline{\rho} }{\partial t }
+ \frac{\partial (\overline{\rho} \; \widetilde{u}_i )}{\partial x_i} = 0  
\end{equation}

\begin{equation}
\frac{\partial  (\overline{\rho} \; \widetilde{u}_i )}{\partial t }+\frac{\partial (\overline{\rho} \; \widetilde{u}_j \;  \widetilde{u}_i )}{\partial x_j } =- \frac{\partial \overline{p}}{\partial x_i } + \frac{\partial (\mu \widetilde{S^d}_{ij} ) }{\partial x_j } -\frac{\partial \tau^{sgs}_{ij}}{\partial x_j} ,
\end{equation}

and

\begin{equation}
\frac{\partial  \overline{E}}{\partial t }+\frac{\partial (\overline{E} \; \widetilde{u}_j)   }{\partial x_j } =- \frac{\partial (\overline{p}\; \widetilde{u_i}) }{\partial x_i } + \frac{\partial   (\mu \widetilde{S^d_{ij}} \widetilde{u_i} )}{\partial x_j } -\frac{\partial (\tau^{sgs}_{ij} \widetilde{u_i})  }{\partial x_j} - \frac{\partial Q_j^{sgs} }{\partial x_j } + \frac{ \partial }{\partial x_j } ( \kappa \frac{\partial \overline{T}}{\partial x_j }) ,
\end{equation}
where $\overline{E} =\overline{\rho}\widetilde{e} + 0.5 \;  \overline{\rho}\widetilde{u_i}\widetilde{u_i}$ is the sum of the resolved internal and kinetic energies. $\widetilde{S^d}_{ij}$ is the deviatoric part of the resolved strain-rate tensor. The relationship between the temperature and the molecular viscosity is assumed to follow a power law with an exponent of 0.75. A constant molecular Prandtl number approximation ($Pr = 0.7$) allows computation of the thermal conductivity. Two additional terms, $\tau^{sgs}_{ij}$ and $Q^{sgs}_{j}$ require modeling closure. The subgrid stress tensor, $\tau^{sgs}_{ij}$ is defined as  $\tau^{sgs}_{ij} = \overline{\rho } (\widetilde{u_i u_j} - \widetilde{u}_i  \widetilde{u}_j) $. Similarly, $Q_j^{sgs} = \overline{\rho} (\widetilde{e u_j} - \widetilde{e}  \widetilde{u_j} ) $ is the subgrid heat flux. In this work, the isotropic component of the subgrid stress is absorbed into pressure, leading to a pseudo-pressure field. The dynamic, tensorial coefficient subgrid-scale model \citep{agrawal2022non} is used in the entirety of this work to model the unresolved scales of motion. Note that the subgrid heat flux is modeled using the constant turbulent Prandtl number approximation ($Pr_t = 0.9$) applied to the dissipative component of the subgrid-stress tensor. %Some more details of the subgrid-scale model are provided below. 

In typical wall-modeled LES, a shear stress, and a heat flux boundary value is supplied to the  LES solver to close the discrete system of governing equations. In this work, an algebraic form of the equilibrium wall-stress model (EQWM) is used, in which the assumed mean velocity profile is a $C^1$ continuous piecewise fit of the viscous sublayer and the logarithmic layer. The details of this wall model can be found in \citet{lehmkuhl2018large}. It is highlighted that a first-point matching approach is used since this solver has not shown any evidence of a ``log-layer mismatch"  \citep{kawai2012wall,yang2017log} in the simulation of turbulent channel flows in the range, $ 1000 \leq Re_{\tau} \leq 4200$ with typical wall-modeled LES resolutions.

\section{A characteristic length scale for separated flows}
The viscously dominated near-wall scales in flows experiencing pressure gradients can go out of equilibrium (for example, not have a linear velocity scaling in viscous units) if the time scale governed by the pressure gradient ($t_p \sim  \rho u_{\tau} / dP/dx $) is faster than the viscous time scale  ($t_v \sim \nu/u_{\tau}^2$) where $u_{\tau}$ is the local skin-friction velocity. The time scale for 
eddies within a logarithmic region is governed by the local shear
($t_{iv} \sim h/u_{\tau}$ where $h$ is the wall-normal height of the flow scale of interest). The inner scales thus go out of equilibrium if 
\begin{equation}
    t_p < t_v  \; \mathrm{or} \;  {u^3_{\tau}} < \frac{\nu}{\rho } \frac{dP}{dx} = (u_p)^3 \implies u_{\tau} < u_p.  
\end{equation}
Similarly, both the inner and outer scales respond at the same time scale if 
\begin{equation}
    t_p < t_v \sim  t_{iv} \; \mathrm{or} \;  
\frac{u_p}{u_{\tau}} > 1 \; \mathrm{and} \;  \frac{u_{\tau} h }{\nu } \sim 1.
\end{equation}

%%%% 
%%%% stb notes: removed ``for an arbitrary h within the boundary 
%%%%% layer''.  the scaling does not hold in the wake region 

\noindent
Near the onset of flow separation, these conditions can be met simultaneously. For a spatially developing boundary layer, these relations also suggest that the flow at a given wall-normal location $h$ is bound to be in non-equilibrium if 
\begin{equation}
   \frac{u_{p} h }{\nu } > 1 \; \mathrm{or} \; \frac{h}{l_p } > 1 \; \mathrm{where} \;  l_p = \frac{\nu}{u_p}. 
\end{equation}

\noindent
Thus, the only scales that remain in equilibrium for a boundary layer before the onset of a separation bubble are scaled such that $\frac{h}{l_p } \sim 1$.  This suggests that an equilibrium approximation of the near-wall flow structure would only remain valid for matching locations at $h \lesssim \alpha l_p$ (for some $\mathcal{O}(1)$ constant, $\alpha$).

%Equivalently, if $h \sim min (\gamma(x) l_p(x))$, where $\gamma(x)$ is some weak function of the streamwise distance that serves as an order unity scaling factor, then for the entirety of the flow, the near-wall motion (at height $h$) can be considered to be in equilibrium. 

% \begin{figure}[!ht] 
%     \includegraphics[width=0.65\textwidth]{laplaceimage.pdf}
%     \caption{Schematic of a simplified, two-dimensional boundary layer flow where $(x_{sep},0)$ is the separation point. The hemisphere of radius $L^m_p$ denotes the region that significantly influences the pressure at the separation point (assumed to be quasi-steady in time).   }
%     \label{fig:laplaceimages}
% \end{figure}

This can be further reasoned from the analysis of a time-averaged Poisson equation for pressure near the separation point, written as 
\begin{equation}
    \langle \rho \frac{\partial \bar{u}_i }{\partial x_j} \frac{\partial \bar{u}_j } {\partial x_i}  \rangle= - \nabla^2 \langle \bar{p} \rangle . 
    \label{eqn:laplace}
\end{equation}
where $\bar{\cdot}$ and $\langle \cdot \rangle$ denote the filtered 
LES fields and a time average operator, respectively, and homogeneous 
Neumann boundary conditions are applied on the pressure field at the solid wall. Consider, a two-dimensional boundary layer that undergoes a separation over a surface such that $\delta/R \ll 1$ where $R$ is the local radius of curvature.  (This assumption may not strictly be true near the separation point over sharp corners, but is convenient for analysis of smooth body separation over gently curved surfaces). Let $(x_{sep}, 0)$ be the coordinates of the separation point. In a semi-infinite domain, the Green's function that satisfies Equation \ref{eqn:laplace} and its boundary conditions is given as

%%%%%%%%%%%
%%%%%%%%%%%

\begin{equation}
    G(x,y, x_0, y_0) = \frac{1}{4 \pi } ln ([(x-x_0)^2 + (y-y_0)^2][(x-x_0)^2 + (y+y_0)^2]) .
\end{equation} for $y \geq 0$.

Using Green's identities and the symmetric property of Green's function, the pressure in the upper-half domain is given as
\begin{equation}
    \langle p^*_{sep} \rangle = \langle p^*(x_{sep},0) \rangle = -  \int^{\infty}_{0} \int^{\infty}_{-\infty} \frac{1}{2 \pi } ln [ (x^*-x^*_{sep})^2 + (y^*)^2 ] \langle  \rho  \frac{\partial u^*_i }{\partial x^*_j} \frac{\partial u^*_j }{\partial x^*_i} \rangle dx^* dy^* .
    \label{eqn:laplacesoln}
\end{equation}
where $(...)^*$ denotes a non-dimensional variable. Let the relevant non-dimensionalization scales be $l_{min} = min (l_p(x)) $ and the viscous velocity scale that corresponds to $l_{min}$, equal to  $u_{max} \approx \nu/l_{min}$. Under this normalization, let $L^{M*}_p$ be the scale of the region that dominantly influences the pressure at the separation point. For this region to be of a finite length scale, the integrand must decrease eventually, or 
\begin{equation}
      ln(L^{M*}_p)^2  \langle \rho \frac{\partial u^*_i }{\partial x^*_j} \frac{\partial u^*_j }{\partial x^*_i} \rangle  \leq \Lambda^2
\end{equation}
for some finite $\Lambda$. For a large-eddy simulation, one of the aims is to find the minimum $L^{M*}_p$ that needs to be resolved on the computational grid to accurately predict separation. The minimum $L^M_p$ that would always satisfy the above inequality is one that also satisfies
\begin{equation}
      (L^{M*}_p)^2   \leq  1 + \frac{\Lambda^2}{max \{ \langle \rho \frac{\partial u^*_i }{\partial x^*_j} \frac{\partial u^*_j }{\partial x^*_i}  \rangle \} }.
\end{equation}
where the logarithm has been Taylor expanded under the expectation that the extent of the dominant upstream region that affects the occurrence of separation is not too dissimilar from the length scale imposed by the strongest adverse pressure gradient.   
\noindent
\citet{stratford1959prediction} suggested that at the points where the skin friction crosses zero, the near-wall velocity profile 
(up to $y_p = \frac{u_{p} y}{\nu } = 5-10$) is given as
\begin{equation}
     \frac{u}{u_p} =   A + B \sqrt{y_p} = A + B \sqrt{\frac{y u_p}{\nu}}.
     \label{eqn:stratfordvel}
 \end{equation}
 
In the context of turbulent flows exhibiting a smooth-body separation due to mild pressure gradients, we assume that the maximum pressure gradient (to which $l_{min}$ corresponds to) is an $\mathcal{O}(1)$ multiple of the pressure gradient at the separation point. This assumption was also numerically verified to hold in the flows considered in the \emph{a-posteriori} sections of this article. Further assuming that near the separation point, the scales of motion in the streamwise and wall-normal directions scale similarly, the continuity equation imposes that the corresponding velocities must also scale similarly.  Then, using Equation \ref{eqn:stratfordvel}, we arrive at 
 \begin{equation}
    \max \{\langle \rho \frac{du^*_i}{dx^*_j}\frac{du^*_j}{dx^*_i} \rangle \} \sim   \mathcal{O}(1).
 \end{equation}
for the near wall region, as the flow nears separation. This implies that
 \begin{equation}
       L^M_p   \sim \chi \; l_{min} \; \mathrm{where} \; \chi \sim  \Lambda .
       \label{eqn:scaling}
 \end{equation}

\noindent
or that a reasonable length scale that may affect a flow nearing separation is the one that is viscously scaled by the strongest adverse pressure gradient in the vicinity of the separation. 

\begin{figure}[!ht]
    \centering{
    \includegraphics[width=0.8 \columnwidth]{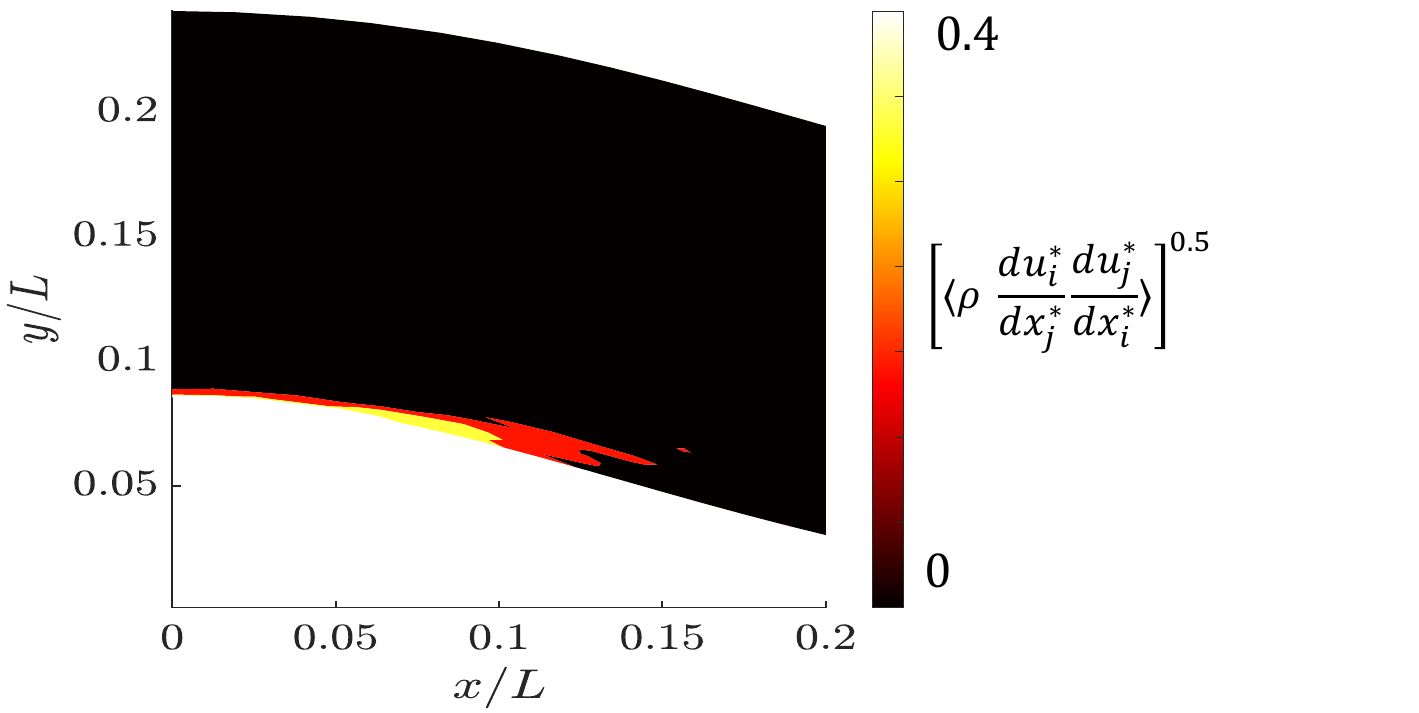} 
    \caption{A two-dimensional (spanwise and time-averaged) contour of $\sqrt{ \langle \rho \frac{du_i}{dx_j} \frac{du_j}{dx_i} \rangle } $ in the vicinity of the separation point ($x/L=0.1$) for the flow over the Boeing speed bump.    }
\label{fig:qcritbump2}    }
\end{figure}

As a numerical example of the analysis above, we consider the flow over the Boeing speed bump \citep{uzun2021high}. A highly resolved wall-modeled LES (with the matching location as low as $\Delta x^+ = \Delta y^+ = \Delta z^+ = 5$ at the wall with an equilibrium wall model) was performed by \citet{whitmore2022brief}. They also showed that the skin friction and the surface pressure coefficients were accurately predicted on this grid in comparison to the quasi-DNS reference data \citep{uzun2021high}. The resulting distribution of the product,  $\sqrt{ \langle \rho \frac{du_i}{dx_j} \frac{du_j}{dx_i} \rangle }$ is plotted in Fig. \ref{fig:qcritbump2}. The maximal value is $\approx 0.4$, implying  $ L^M_p   \sim  \mathcal{O}(1)  $.  

The value of $\chi$ may vary depending on the flow; however, this will not affect the scaling $L^M_p   \sim l_{min} $. Some reasonable estimates of $\Lambda$ are explored through \emph{a-priori} analysis in the next subsection. A further implication of this fact can be that the response of the near-wall flow in the vicinity of a separation bubble can also be governed by a stronger adverse pressure gradient upstream (locally close to the point of separation, however) of this region. 
%Indirectly, performing wall-modeled LES by forcing the matching location at the proposed global length scale suggests that LES will resolve all the history effects so that the wall model will not have to address them. 

%%%
%% stb comment for section 3.1: 
%%% removed An added emphasis is put on the monotonic grid-convergence owing to the \emph{a-posteriori} performance of wall-modeled LES in certain mild pressure-gradient regimes \citep{agrawal2022non,whitmore2022brief}.
%%% why bring up the non-monotonicity here? 
%%%%%

\subsection{\emph{A-priori} analysis}
%\vspace{-5pt}
In this section, \emph{a-priori} analysis is performed to obtain an estimate for the expected values of $\Lambda$ in non-equilibrium boundary layers. It is highlighted that a value of $\Lambda \leq 1$ does not necessarily imply the limit of ``wall resolved LES", however, for practical wall-modeled LES, $\Lambda \geq 1$ is useful. Hence, for this analysis, the objective is to find reasonable values of $\Lambda$ that provide monotonically converging, and fairly accurate (for example, $\leq 10-15\% $ error) estimates of the wall shear stress for non-equilibrium flows. The equilibrium wall model is used for providing the estimates of the modeled wall stress. For this purpose, a high-fidelity database consisting of the adverse pressure gradient boundary layers of \citet{bobke2017history} and the flow over NACA airfoils \citep{tanarro2020effect,vinuesa2018turbulent} is examined. A momentum-thickness-based Reynolds number ($Re_{\theta}$) up to approximately 4000 and a Clauser parameter value ($\beta =  \delta^* /\rho u^2_{\tau} \; dP/dx $) of up to 6 is observed in these flows. 

\begin{figure}[!ht]
    \centering{
    \includegraphics[width=1\columnwidth]{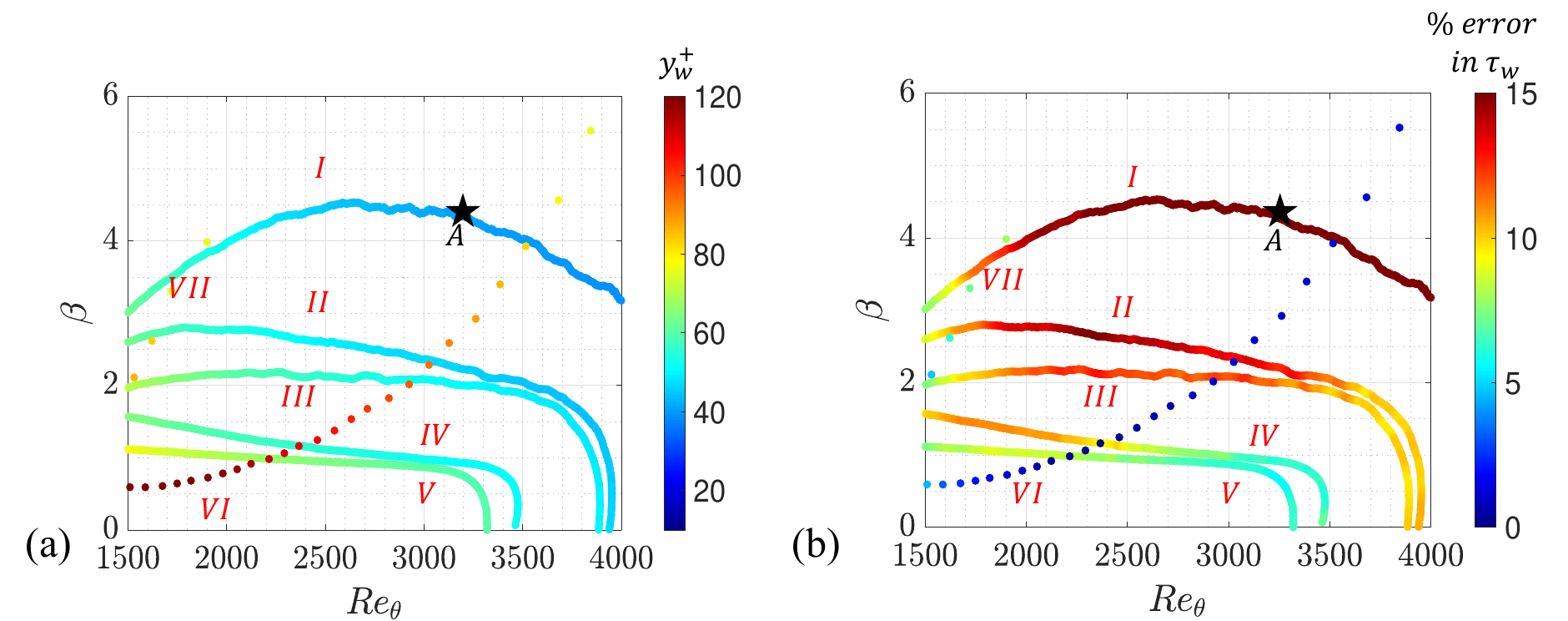}}
    \caption{The distribution of the (a) location of the matching point in wall units ($y^+_w$) and (b) the corresponding percentage error in the prediction of the wall shear stress from the equilibrium wall model for $\chi=25$ as a function of the Clauser parameter and the momentum thickness Reynolds numbers. The Roman numerals denote the data points from different flows: Data series I to V correspond to different streamwise stations in the flat-plate boundary layers simulated in \citet{bobke2017history}. Data series VI and VII correspond to different streamwise stations in NACA 0012 and 4412 airfoils studied by \citet{tanarro2020effect} and \citet{vinuesa2018turbulent} respectively. The starred point, A, is a sample point chosen to aid the interpretability of this plot. This data point corresponds to a flat-plate boundary layer flow at $Re_{\theta} \sim 3200 \; \mathrm{and} \; \beta=4.2$. The matching location for this point is roughly, $y_w^+ \approx 40$ [in subfigure (a)] and the percentage error in the shear-stress prediction is approximately 20\% [and hence the contour is saturated in subfigure (b)].  }
    \label{fig:apriori24}
\end{figure}

\begin{figure}[!ht]
    \centering{
    \includegraphics[width=1\columnwidth]{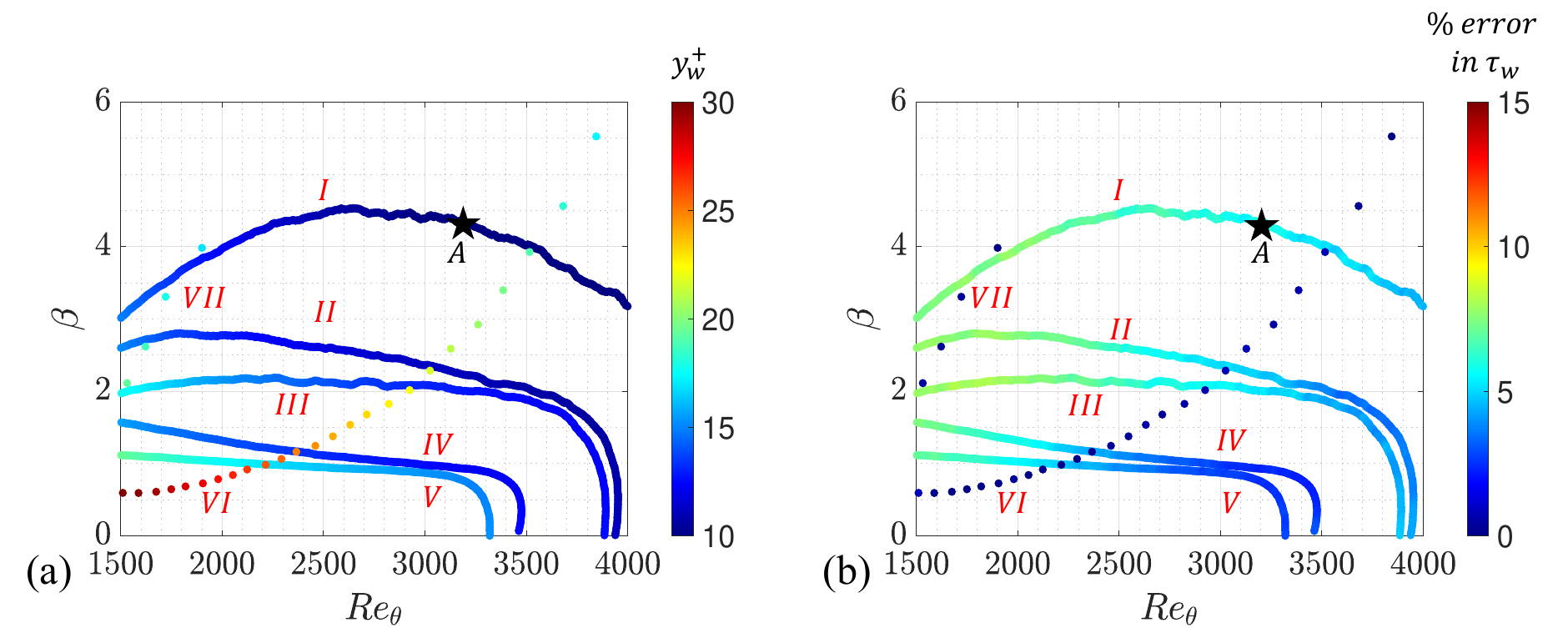}}
    \caption{The distribution of the (a) location of the matching point in wall units ($y^+_w$) and (b) the corresponding percentage error in the prediction of the wall shear stress from the equilibrium wall model for $\chi=25$ as a function of the Clauser parameter and the momentum thickness Reynolds numbers. The Roman numerals denote the data points from different flows: Data series I to V correspond to different streamwise stations in the flat-plate boundary layers simulated in \citet{bobke2017history}. Data series VI and VII correspond to different streamwise stations in NACA 0012 and 4412 airfoils studied by \citet{tanarro2020effect} and \citet{vinuesa2018turbulent} respectively. The starred point, A, is a sample point chosen to aid the interpretability of this plot. This data point corresponds to a flat-plate boundary layer flow at $Re_{\theta} \sim 3200 \; \mathrm{and} \; \beta=4.2$. The matching location for this point is roughly, $y_w^+ \approx 10$ [in subfigure (a)] and the percentage error in the shear-stress prediction is approximately 6\% [in subfigure (b)].   }
    \label{fig:apriori6}
\end{figure}

In Figures \ref{fig:apriori24} and \ref{fig:apriori6}, the value of $\chi$ decreased from 25 to 6. For both these figures, subfigure (a) denotes the location of the matching point ($y^+_w$) in wall units with the ordinate and the abscissa denoting the local Clauser parameter and the momentum thickness-based Reynolds number respectively. Similarly, subfigure (b) presents the percentage error in the prediction of the shear stress using the equilibrium wall model. As an example, the starred point, A, in Fig. \ref{fig:apriori24} corresponds to a flat-plate boundary layer flow at $Re_{\theta} \sim 3200 \; \mathrm{and} \; \beta=4.2$ with $y_w^+ \approx 40$ (in Fig. \ref{fig:apriori24}(a)) and the percentage error in the shear-stress prediction is approximately 20\% (in Fig. \ref{fig:apriori24}(b)). Generally, in Fig. \ref{fig:apriori24}, the error in the prediction of the shear stress, $\tau_w$, reaches up to approximately 30\% with the error generally increasing for a higher $y^+_w$ except when the Reynolds number is low. At a given $Re_{\theta}$, the error in the shear-stress predictions increases with increasing $\beta$. This is expected as for a larger $\beta$, a large range of the flow scales within the boundary layer are under non-equilibrium. A turbulent channel flow analog of this argument can be found in \citet{lozano2020non}. At a given $\beta$, the increase in $Re_{\theta}$ generally leads to a lower error as the scale separation between the equilibrium and non-equilibrium regions of the boundary layer increases. Upon refinement in Fig. \ref{fig:apriori6}, by reducing $\chi$ to 6 (equivalent to quadrupling the resolution), the errors are lower than those at $\chi = 25$. The errors for all the streamwise stations in the considered flows are less than $10\%$ in this limit. Given the relatively modest $Re_{\theta}$ of these flows, the corresponding $y^+_w$ values are also small. However, based on these \emph{a-priori} results, it is hypothesized that a value of $\chi \in [5,10]$ may be necessary to accurately predict quantities of interest in wall-modeled LES of non-equilibrium flows, atleast while leveraging the equilibrium wall model.

%\vspace{-20pt}
\subsection{Reynolds number scaling}
%\vspace{-5pt}
The currently accepted zero-pressure-gradient flow-based estimates \citep{choi2012grid,yang2021grid} suggest that the number of grid points ($N_{cv}$) required to perform wall-modeled LES of a boundary layer scale as $N_{cv} \sim Re^1$. Instead, in this work, we propose an alternate grid-point scaling for non-equilibrium flows, specifically those that exhibit a turbulent separation.  These estimates 
will be tested in \emph{a-posteriori} calculations in subsequent 
sections. The assumption in this work is that a nearly constant value of $\chi$, or a viscously driven length scale, can be chosen to predict the quantities of interest as the Reynolds number is increased. 
The value of $L^M_{p}$ scales as 
\begin{equation}
    L^M_{p} \sim \chi l_{min} \sim \chi Re^{-2/3}  \max\left(\frac{1}{\rho}\frac{dP}{dx}\right)^{-1/3}.
\end{equation}
or that as the Reynolds number is increased, $L^M_{p}$ decreases as $Re^{-2/3}$. Thus, for a nested grid where the number of points in the wall-normal direction with respect to $L^M_p$ is fixed, the number of grid points required to sufficiently resolve the non-equilibrium effects scales as $N_{cv} \sim Re^{4/3}$. This prediction is more stringent than the equilibrium flat-plate boundary layer wall-modeled LES estimates but grows slower than $N_{cv} \sim Re^{13/7}$ scaling of wall-resolved LES. The assumption of requiring a fixed $\chi$ across a range of Reynolds numbers may be negated once improved wall models consistently and accurately account for the non-equilibrium and the flow history effects at coarse resolutions. In Appendix I, the performance of some existing non-equilibrium wall models is reported, which 
suggests that present non-equilibrium models do not affect the resolution requirements significantly. Following the works of \citet{chapman1979computational,choi2012grid,yang2021grid} Appendix II also provides some \emph{a-priori} estimates of the number of points required for wall-modeled LES of non-equilibrium boundary layers assuming a fixed outer resolution ($\Delta/\delta = \mathrm{const.}$). However, it is noted that these \emph{a-priori} estimates may not be realized in \emph{a-posteriori} simulations owing to the errors in the subgrid-scale and wall models. 

\section{\emph{A-posteriori} analysis}
%\vspace{-5pt}
\subsection{Wall-modeled LES of the Boeing speed bump}
%\vspace{-5pt}

\label{sec:bump}
In this section, \emph{a-posteriori} results from wall-modeled LES of the flow over the Boeing speed bump \citep{williams2020experimental,gray2021new,gray2022experimental} are presented. This geometry was proposed by the Boeing Company and \citet{williams2020experimental} for validation of CFD methods in the prediction of turbulent flow separation. 
\citet{zhou2023large} recently showed a large sensitivity in the prediction of the separation bubble to the choice of the subgrid-scale and wall models. The simplified, spanwise-periodic bump surface is defined by an analytical expression, $h(x)$, written as
\begin{equation}
    y\left(\frac{x}{L}\right) = 0.085 \, exp\left[-\left(\frac{x}{0.195L}\right)^2\right].
    \label{eqn:bump}
\end{equation}

\begin{figure}[!ht]
    \centering
    \includegraphics[width=0.8\columnwidth]{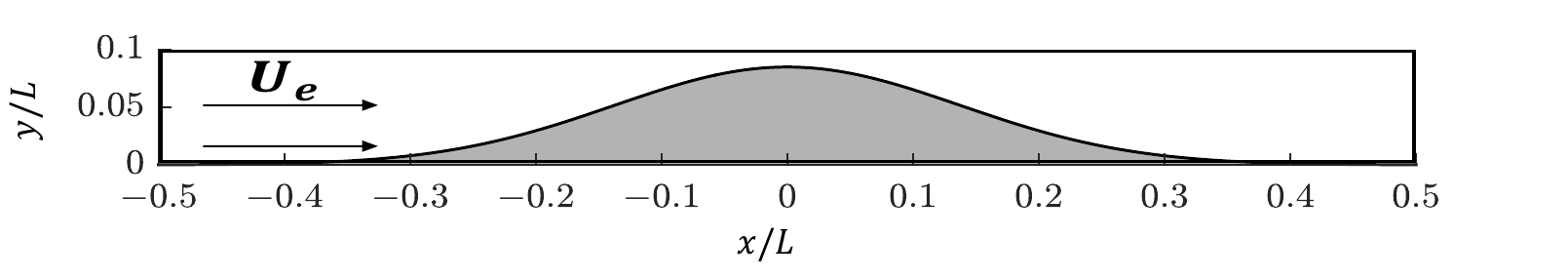}
    \caption{A schematic of the flow over the simplified, spanwise periodic Boeing speed bump geometry. The flow develops over a Gaussian-shaped bump experiencing a favorable and an adverse pressure gradient before eventually separating at $x/L \approx 0.1$.  }
    \label{fig:speedbumpgeom}
\end{figure}

\noindent
where $x$ is the streamwise coordinate. The Reynolds number, $Re_L$ for this flow is defined in terms of the freestream velocity $U_{\infty}$ and the bump width $L$.  It has been previously shown that the surface pressure ($C_p$) and the skin friction ($C_f$) for the simplified geometry approximately match those at the midspan of the experimental configuration \citep{gray2021new}. In this work, we first perform computations at a Reynolds number, $Re_L = 2 \times 10^6$, to match the prior quasi-DNS \citep{uzun2021high} and for comparison with experiments \citep{williams2020experimental,gray2021new} at $Re_L = 3.41 \times 10^6$ and $Re_L = 4 \times 10^6$. The flow over the speed bump experiences a strong favorable pressure gradient followed by a strong adverse pressure gradient on the fore and aft sections of the bump respectively. For $Re_L \geq 1.8 \times 10^6$, these pressure gradients lead to the formation of a turbulent separation bubble.

\begin{table}
\centering
\begin{tabular}{ p{1.5cm}p{1.5cm}p{1.5cm}p{1.5cm} }

Mesh & $N_{cv}$ & max $\Delta / L$  & min $\Delta / L$ \\
\hline\noalign{\vspace{3pt}}
Coarse  & $3$ Mil. & $0.01$ & $1.3 \times 10^{-3}$ \\
Medium  & $12$ Mil. & $0.01$ & $6.3 \times 10^{-4}$ \\
Fine    & $52$ Mil. & $0.01$ & $3.1 \times 10^{-4}$ \\
%Very Fine    & $200$ Mil. & $0.01$ & $1.5 \times 10^{-4}$ \\
\end{tabular}
\caption{Mesh parameters for the $Re_L = 2  \times 10^6$ case of the flow over the spanwise-periodic Boeing speed bump. For subsequently higher Reynolds numbers, the grid is refined by modifying the background resolution (equal to max. $\Delta/L$) by a factor of $Re^{2/3}$.  }
\label{table:resbump2}
\end{table}

Fig. \ref{fig:speedbumpgeom} represents a schematic of the simulation setup. The inlet is located at $x/L = -1.0$ as a plug-flow profile, and the flow then undergoes a transition to turbulence over the region, $-1.0 < x/L < -0.6$. This simple choice stems from the observations of \citet{agrawal2023thwaites} who showed that the separation tendency of this flow is approximately independent of the inlet boundary condition as long as the inlet is sufficiently far upstream. This has also been reported in \emph{a-posteriori} simulations by \citet{whitmore2022brief}. 
Free-stream velocities with a non-reflective boundary condition are set at the top boundary. A characteristic boundary condition (NSCBC) with constant pressure is applied at the outlet ($x/L = 2.5$). The computational meshing approaches are the same as in \citet{agrawal2022non}; some details are provided in Table \ref{table:resbump2}. A schematic of the computational mesh as a function of the streamwise coordinate is provided in Appendix III.

The experiments of \citet{williams2020experimental} and \citet{gray2021new} have suggested an approximate Reynolds number independence in the surface pressure distribution over the bump surface for $Re_L \geq 2 \times 10^6$. In light of this, in this work, we perform numerical experiments up to $Re_L = 10 \times 10^6 $, assuming the invariance of the pressure coefficient profile holds at the increased Reynolds numbers. Only 
comparisons of the surface pressure distribution are made as it 
is expected that the skin-friction distribution in the upstream region (nearly a zero pressure gradient flat-plate flow) is expected to be a function of the Reynolds number. In a prior work \citep{agrawal2022non}, the authors showed that for the Reynolds number, $Re_L = 2 \times 10^6$ case, the skin friction compares well with the experiments. The surface pressure coefficient ($C_p$) is mathematically defined as 
\begin{equation}
   \frac{C_p}{2} = \frac{ p - p_{\rm ref}}{ \rho_{\infty} U^2_{\infty}}.
\end{equation}
where $U_{\infty}$, $p$, and $p_{\rm ref}$ are the mean free-stream velocity, wall pressure, and the upstream reference pressure in the zero-pressure region of the flow respectively.

%%%
%%% stb notes: figure 6d needs data from all the Re, not just the 
%%% finest grid level.  the y-axis for figure 6d should start at 
%%% 0.  lastly, the inclusion of the factor of 2 on the x-axis is 
%%% confusing.  if you are trying to correct for the half-height, 
%%% just call it \Delta as a nominal measure of the grid length scale.
%%%% 

\begin{figure}[!ht]
    \centering
    \includegraphics[width=1\columnwidth]{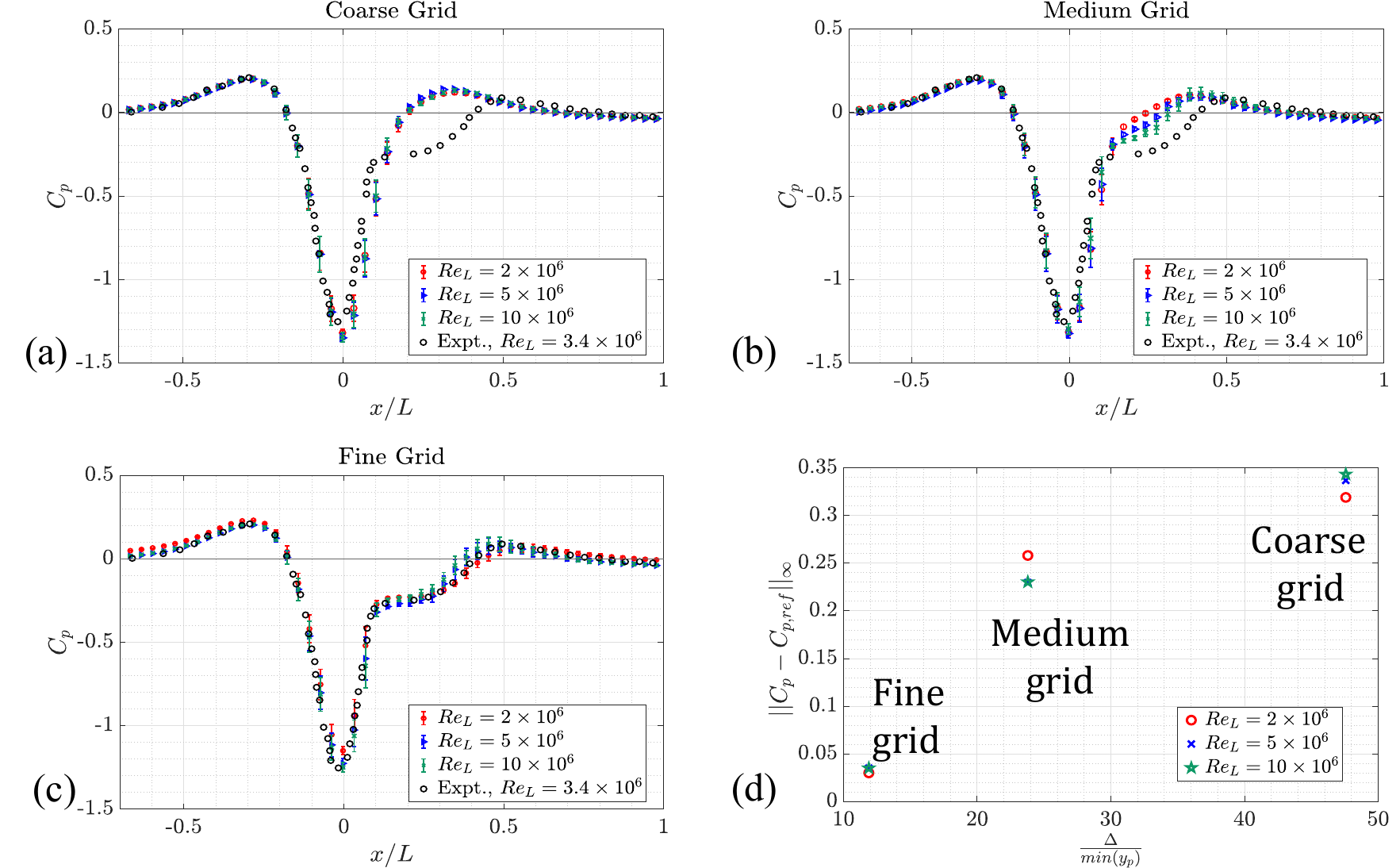}
    \caption{The streamwise distribution of the surface pressure coefficient, $C_p$ for the (a) coarse grid, (b) medium grid, and (c) fine grid across $2 \times 10^6 \leq Re_L \leq 10 \times 10^6$. Note that across different Reynolds numbers, the nominal ``coarse", ``medium" and ``fine" grids are scaled in their resolution by a factor of $Re_L^{2/3}$.  Subfigure (d) shows the $L_{\infty}$ norm of the error in $C_p$ for all the Reynolds number flows. The vertical bars in the simulations in subfigures (a)-(c) represent the root mean square error around the mean value of $C_p$. }
    \label{fig:speedbumpcp}
\end{figure}

In Fig. \ref{fig:speedbumpcp}, the predictions of the surface pressure as a function of the Reynolds number are plotted. Subfigures (a)-(c) provide these predictions at a given nominal grid for different $Re_L$, 
where at each level (e.g., coarse, medium, fine), grids are refined based on the proposed Reynolds number scaling ($Re^{2/3}_L$).  When the resolution of the calculation with respect to $l_p$ is held constant, it is apparent that predictions of wall-modeled LES at different Reynolds numbers are similar. For all the considered cases, the finest grid is required to accurately predict $C_p$, specifically, the flow separation ($0.1 \leq x/L \leq 0.4$). To quantify these errors, Fig. \ref{fig:speedbumpcp}(d) presents the $L_{\infty}$ norm of the error in $C_p$. This norm is chosen in the entirety of this work to specifically highlight the largest differences in wall-modeled LES and reference values; which occur primarily due to incorrect prediction of the extent of the separation bubble (since the distribution of $C_p$ sufficiently far upstream and downstream of the separation bubble can be described mostly from inviscid flow arguments). The error values are expected to reach an asymptotic value on very coarse grids when the prediction of $C_p$ reaches the inviscid limit (the flow remains attached and the boundary layers are sufficiently thin). The error convergence on the fine grid corresponds to a $\Delta/ min(l_p) \sim 12 $, which is comparable to the values from the \emph{a-priori} analysis in the preceding sections. This is also encouraging because the preceding analysis does not account for numerical and subgrid-scale modeling errors. Appendix IV provides a discussion on the limited sensitivity in the convergence behavior of wall-modeled LES on the choice of the subgrid-scale model between the dynamic Smagorinksy model \citep{moin1991dynamic} and the dynamic tensor coefficient Smagorinsky model \citep{agrawal2022non}. Although not shown, it was observed that the minimum $y^+$ values before the separation point are approximately $ y^{+}_w \approx 15 - 20$, implying the simulations are not wall-resolved. An additional simulation on a ``very-fine" grid (twice as refined in each direction than the ``fine'' grid, not shown) for the $Re_L = 5 \times 10^6$ case confirmed that the results are grid-converged.

\subsection{Wall-modeled LES of the Song-Eaton Diffuser}
%\vspace{-5pt}

\begin{figure}[!ht]
    \centering
    \includegraphics[width=0.85\columnwidth]{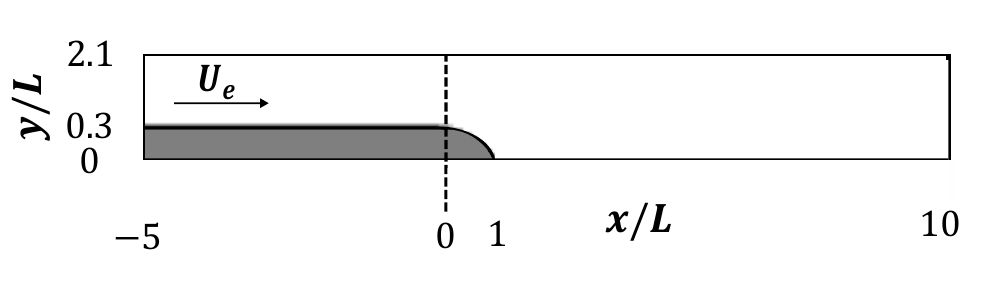}
    \caption{A schematic of the simulation setup for the flow over the arc-shaped diffuser studied by \citet{song2004reynolds}. The flow develops over a flat plate upstream of the arc before experiencing an adverse pressure gradient that leads to a flow separation at $x/L \approx 0.79$. The vertical dashed line at $x/L=0$ marks the beginning of the diffuser's arc.   }
    \label{fig:songeatongeom}
\end{figure}

\citet{song2004reynolds} performed detailed experiments on the Reynolds number sensitivities of flow separation over a circular arc-shaped diffuser. The experiments recorded a small separation bubble with nearly Reynolds number invariant $C_p$ distribution across a range of $ 3400 \leq Re^{ref}_{\theta} \leq 20100$ where $Re^{ref}_{\theta}$ denotes the momentum thickness based Reynolds number at a reference station ($x/L = -2 $ where $L$ is the length of the arc of the diffuser) in the flat-plate region of the diffuser. Fig. \ref{fig:songeatongeom} presents a schematic of the simulation setup. In this flow, the unsteady separation process occurs over a large portion of the arc surface leading to the formation of intermittently attached and separated boundary layers. \citet{radhakrishnan2006large} performed a detached-eddy simulation of this flow with a stochastic forcing \citep{keating2006dynamic} in the ``transition-region" between quasi-steady RANS near-wall flow to the unsteady LES outer flow; this results in an \emph{ad hoc} breakdown of the streamwise coherence of the eddies. Their results reasonably captured the separation bubble for the $Re^{ref}_{\theta} = 13200$ flow, however, they employed the conventional RANS type stretched grids in the wall-normal direction. The near-wall cells had an approximate aspect ratio of  $300:1$ and were refined down to $y^+ = 1$.

Table \ref{table:songeatontable} presents details of the computational grids in this work. A schematic of the computational mesh as a function of the streamwise coordinate can be found in Appendix III. The simulation setup for this flow is very similar to the Boeing speed bump. A plug-flow is fed at the inlet ($x/L=-5$) which transitions into an equilibrium turbulent boundary layer, as it develops over a long flat-plate region before encountering the diffuser. The outlet boundary in the simulations was extended up to $x/L=10$ to let the flow fully recover post-reattachment. The top boundary is placed at $y/L = 2.1$; both the top and the bottom walls are treated using the equilibrium wall model. The outlet is treated with the characteristic boundary condition \citep{poinsot1992boundary}. Unlike \citet{radhakrishnan2006large}, the grid resolution is isotropic in all three spatial directions near the wall. Periodic boundary conditions are used along the spanwise direction with a period of $10 \; \delta^{ref}_{99}$ where $\delta^{ref}_{99}$ is the experimental boundary layer thickness at the reference upstream station ($x/L=-2$).

\begin{table}
\centering
\begin{tabular}{ p{1.5cm}p{1.5cm}p{1.5cm}p{1.5cm} }

Mesh & $N_{cv}$ & max. $\Delta / L$ & min. $\Delta / L$   \\
\hline\noalign{\vspace{3pt}}
Coarse  & $0.4$ Mil. & $0.03$ & $7.6 \times 10^{-3}$ \\
Medium  & $1.3$ Mil. & $0.03$ & $3.8 \times 10^{-3}$ \\
Fine    & $5.2$ Mil. & $0.03$ & $1.9 \times 10^{-3}$ \\
\end{tabular}
\caption{Mesh parameters for the Reynolds number, $Re^{ref}_{\theta} = 3400$ (based on the momentum thickness at the upstream station, $x/L=-2$) case of the flow over the circular arc-shaped diffuser studied by  \citet{song2004reynolds}. For subsequently higher Reynolds numbers, the grid is refined by modifying the background resolution (equal to max. $\Delta/L$) by a factor of $Re^{2/3}$.  }
\label{table:songeatontable}
\end{table}

\begin{figure}[!ht]
    \centering
    \includegraphics[width=1\columnwidth]{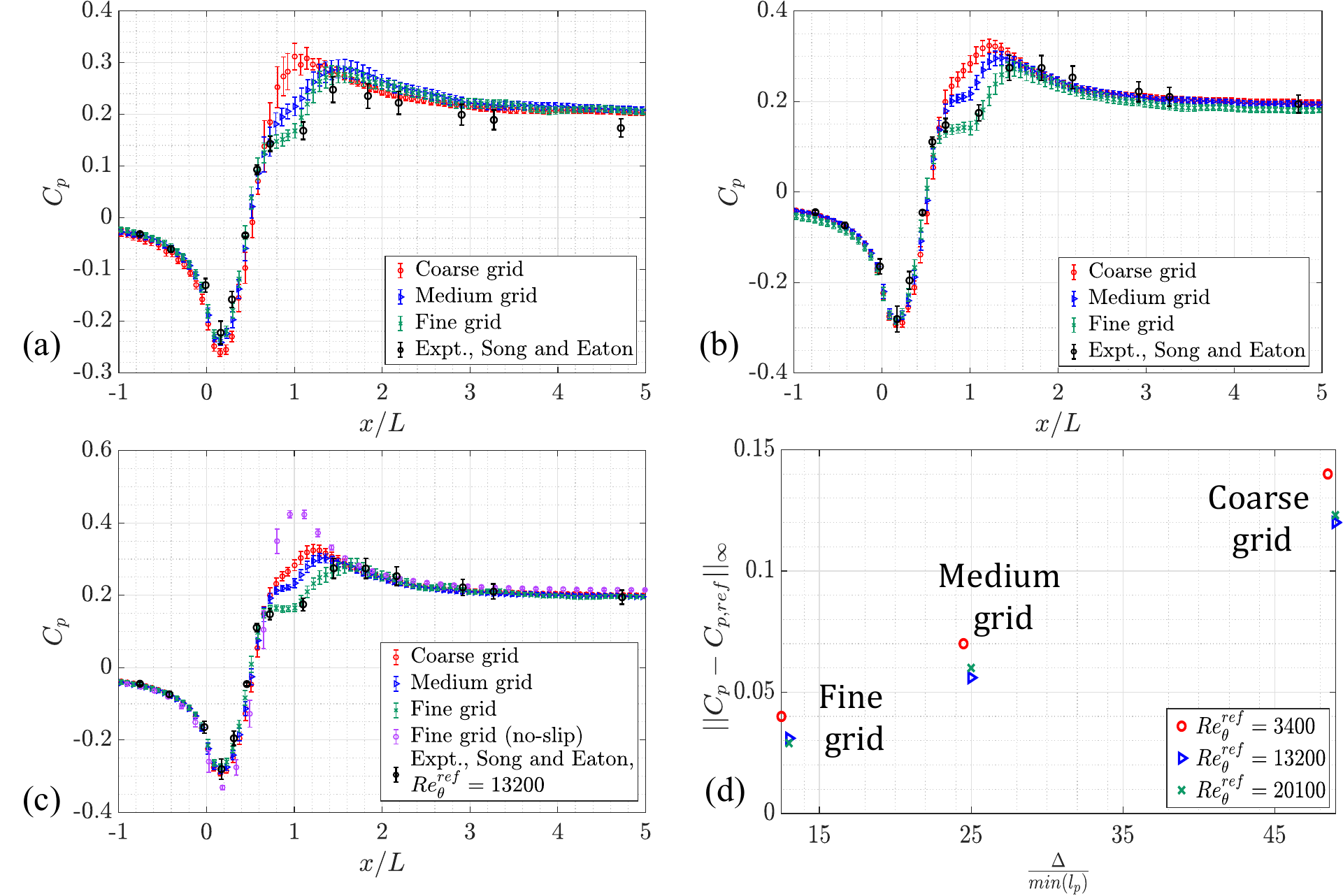}
    \caption{The prediction of surface pressure coefficient $C_p$ for the flow over the Song-Eaton diffuser at (a) $Re^{ref}_{\theta}=3400$, (b) $Re^{ref}_{\theta}=13200$, and (c) $Re^{ref}_{\theta}=20100$ respectively. The flow separates at $x/L \approx 0.79$ and reattaches at $x/L \approx 1.2$. Subfigure (d) provides the error convergence plot upon grid-refinement for different Reynolds numbers. The vertical bars in the simulations and the experiment in subfigures (a)-(c) represent the uncertainty in the mean values. In the simulations, the root mean square error was used to quantify this uncertainty. The reference $C_p$ in $Re^{ref}_{\theta}=20100$ simulations is taken from $Re^{ref}_{\theta}=13200$ experiments due to lack of the availability of experimental data at $Re^{ref}_{\theta}=20100$ in \citet{song2004reynolds}.   }
    \label{fig:songeatoncp}
\end{figure}

%%%
%%% stb notes: 
%%% excerpted from the paragraph below: Although not shown, at $Re_{\theta}=20100$, a fine grid simulation with no-slip boundary conditions on the bottom walls produces the inviscid $C_p$ solution and does not predict any flow separation which confirms the necessity of a wall model to simulate this flow at the finest grid resolution.
%%% 
%%% i think that you should show this.  this reinforces that you 
%%%% are not converging because you have reached wall resolved 
%%% resolutions 

\noindent
We performed wall-modeled LES of all three Reynolds numbers, $Re^{ref}_{\theta} = 3400, \; 13200, \; 20100$ considered in the experiments. Fig. \ref{fig:songeatoncp} shows the $C_p$ distribution for these cases. Subfigures (a)-(c) correspond to the three Reynolds numbers respectively. The $C_p$ distribution is nearly identical for all Reynolds numbers, with the flow experiencing an acceleration slightly upstream of the arc (as $x/L \rightarrow 0$). Then, as the boundary layer experiences an adverse pressure gradient over the arc, a small separation occurs at $x/L \approx 0.79$. For all Reynolds numbers, the fine grid accurately captures the separation bubble and the flow recovery regions. Fig. \ref{fig:songeatoncp} also shows that at $Re_{\theta}=20100$, a fine grid simulation with no-slip boundary conditions on the bottom walls produces the inviscid $C_p$ solution and does not predict any flow separation which confirms the necessity of a wall model to simulate this flow at the finest grid resolution. For this flow, the performance of wall-modeled LES for coarse and medium grids varies between the lowest and the other two Reynolds numbers similar to the Boeing speed bump. Fig. \ref{fig:songeatoncp}(d) presents the $L_{\infty}$ norm of the error in the $C_p$; with the errors becoming ``small" only for the finest grid. Selected sensitivity tests (not shown) to the choice of the subgrid-scale model suggested no differences in the grid-point scaling with the Reynolds number.  

\subsection{Wall-modeled LES of the Notre-Dame Ramp }
%\vspace{-5pt}
\begin{figure}[!ht]
    \centering
    \includegraphics[width=0.9\columnwidth]{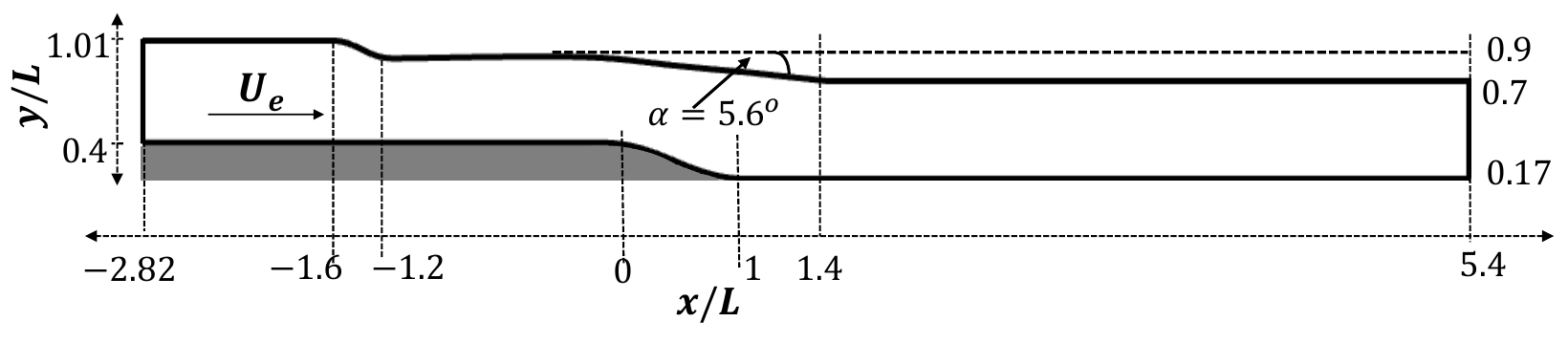}
    \caption{A schematic of the flow over the smooth ramp studied by \citet{simmons2020experimental}. The flow develops over a flat plate upstream of the ramp before separating at $x/L \approx 0.5$. The dotted lines are only plotted for marking the reference coordinates and are not a part of the simulation domain. The curved portion of the upper surface in the upstream region ($x/L < 0$) is replicated from the experimental geometry to provide a weak favorable pressure gradient to accelerate the incoming flow.   }
    \label{fig:notredamegeom}
\end{figure}

In this section, we further test the proposed $\Delta/min(l_p) \sim \mathcal{O}(10) $ resolution requirements for predicting turbulent separation accurately on the flow over a smoothly curved ramp. \citet{simmons2020experimental} performed a series of experiments on a smooth ramp over which a high Reynolds number boundary layer (upstream reference $Re_{\theta} = 11000$ at $x/L=-0.75$, or, $Re_H = U_{\infty} H^{r} / \nu = 8.4 \times 10^5 $ where $H^r$ is the height of the ramp) exhibits a smooth-body separation; this flow is a candidate for the ``High-Fidelity CFD Verification"  Workshop at AIAA SciTech, 2024. The top wall of the experimental setup was slightly tapered to create a pressure gradient, with the tapering angle having a significant impact on the extent of the separation \citep{simmons2020experimental}. Fig. \ref{fig:notredamegeom} provides a schematic of the simulation setup that nearly reproduces the experimental wind tunnel configuration. The ramp geometry ($0 < x/L < 1$ on the bottom wall) is obtained from a fifth-order polynomial that smoothly joins the upstream and downstream flat plates. Three tapering angles, $\alpha=3.2^o, \; 5.6^o, \; 7.7^o$ were studied experimentally, with the smallest angle producing the largest separation bubble, and the largest angle producing a fully attached flow. In this work, we consider the $\alpha=5.6^o$, expecting that this angle will be the most challenging condition in regard to the prediction of the separation bubble. 

\begin{table}
\centering
\begin{tabular}{ p{2cm}p{2cm}p{2cm}p{2cm}p{2cm} }
Mesh & $N_{cv}$ & max. $\Delta / L$ & min. $\Delta / L$   \\
\hline\noalign{\vspace{3pt}}
Coarse  & $16$ Mil. & $0.02$ & $2.5 \times 10^{-3}$  \\
Medium  & $64$ Mil. & $0.02$ & $1.3 \times 10^{-3}$   \\
Fine    & $252$ Mil. & $0.02$ & $6.25 \times 10^{-4}$  \\
\end{tabular}
\caption{Mesh parameters for the $Re^{ref}_{\theta} = 11000$ (Reynolds number based on the momentum thickness at $x/L=-0.75$) or $Re_H = 8.4 \times 10^5$ (based on ramp-height) case of the flow over the smooth ramp studied by \citet{simmons2020experimental}.  }
\label{table:notredametable}
\end{table}

In our computations, the inlet is fed as a plug flow located at $x/L = -2.82$. To approximately match the experimental boundary conditions, both the top and the bottom walls are treated inviscidly in the region $-2.82 \leq x/L \leq -1.6$. Downstream of this region, the bottom wall flat plate and the ramp are treated with the equilibrium wall model. The top wall is located at the same height as in the experiments, the equilibrium wall model closure is also used at this boundary. The smooth curvature of the top wall (between $-1.6 \leq x/L \leq -1.2$ and $0 \leq x/L \leq 1.4$) is replicated from the experiments to provide a weak favorable pressure gradient to the flow. In the experiments, after the flow recovers from the separation, it leaves the test section at $x/L \approx 1.4$ to enter a diffuser region. In our simulations, the top wall is flattened after $x/L = 1.4$ and the flow is allowed to recover downstream in a nominally zero-pressure gradient region with the outlet located at $x/L=5.4$. The outlet boundary condition is a non-reflective characteristic boundary condition \citep{poinsot1992boundary}. The domain is $\approx 25 \delta^{ref}_{99}$ wide in the spanwise direction. Details of the grid distribution and resolutions are provided in Table \ref{table:notredametable}. The schematics of the computational mesh are provided in Appendix III.

Fig. \ref{fig:simmonscp}(a) shows the $C_p$ distribution across the grid refinement sweep. As the flow encounters the ramp, a favorable pressure gradient is experienced initially due to the tapering of the top wall that creates a flow acceleration. However, at $x/L \approx 0.2$, the adverse pressure gradient due to the ramp surface dominates and leads to the formation of a separation bubble at $x/L \approx 0.5$. In the current wall-modeled LES results, on the coarsest grid, the flow does not separate. As the grid is refined, the flow starts separating, and on the fine grid, the solution is nearly identical to the experimental measurement of $C_p$. Table \ref{table:notredametable} and Fig. \ref{fig:simmonscp}(b) suggest that for the fine grid, the resolution in terms of the pressure-scaled units is approximately the same as that required on the Boeing speed bump, and the Song-Eaton diffuser flow.

\begin{figure}[!ht]
    \centering
    \includegraphics[width=1\columnwidth]{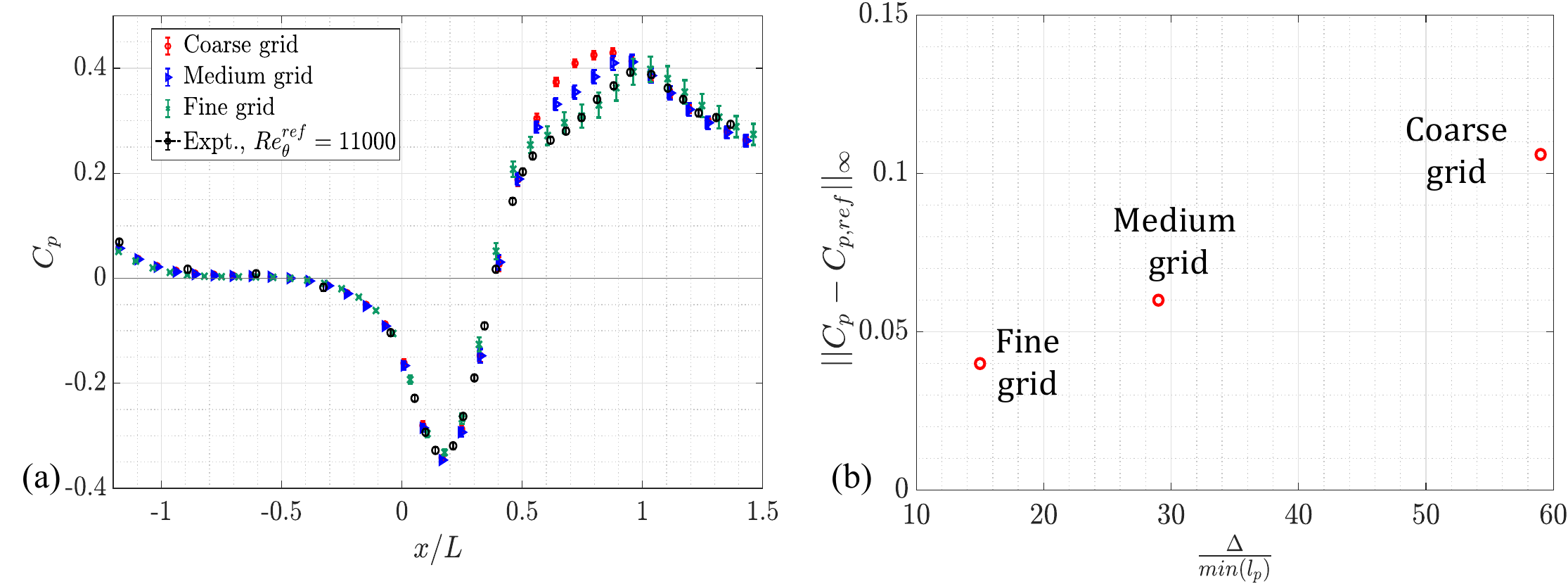}
    \caption{The (a) surface pressure distribution and the (b) error convergence upon grid-refinement plots for the flow over the smooth-ramp designed and studied by \citet{simmons2020experimental}. 
    The flow develops over a flat plate upstream of the ramp before separating at $x/L \approx 0.5$, and later reattaching with the bottom wall at $x/L \approx 0.90$. The vertical bars in the experiment in subfigure (a) represent the uncertainty in the mean values.}
    \label{fig:simmonscp}
\end{figure}

\subsection{Wall-modeled LES of the Backward-Facing step}
%\vspace{-5pt}
The flow over the backward-facing step has been extensively studied, both experimentally \citep{jovic1994backward,driver1985features,adams1988flow} and computationally \citep{le1997direct,akselvoll1993large,pont2019direct} to describe the structure of a turbulent separating boundary layer. \citet{jovic1994backward} observed that the reattachment length of the separated flow past the step is a weak function of the Reynolds number. The separation over the step is a geometric constraint in this flow, but the prediction of the reattachment length is dependent on the induced pressure gradient due to the extent of the separation bubble. Unlike the other cases considered in this work, the flow past the separation point for the backward-facing step experiences the strongest pressure gradient post-separation. It is highlighted that given its geometrically imposed separation point, this flow allows solely focusing on examining the resolution requirements for accurate prediction of the reattachment point.
\citet{jovic1994backward} also reported a weak Reynolds number dependence of the separated flow for different expansion ratios, $E_r$ (equal to the ratio of the height of the region downstream of the step to the region upstream of it).

\begin{figure}[!ht]
    \centering
    \includegraphics[width=1\columnwidth]{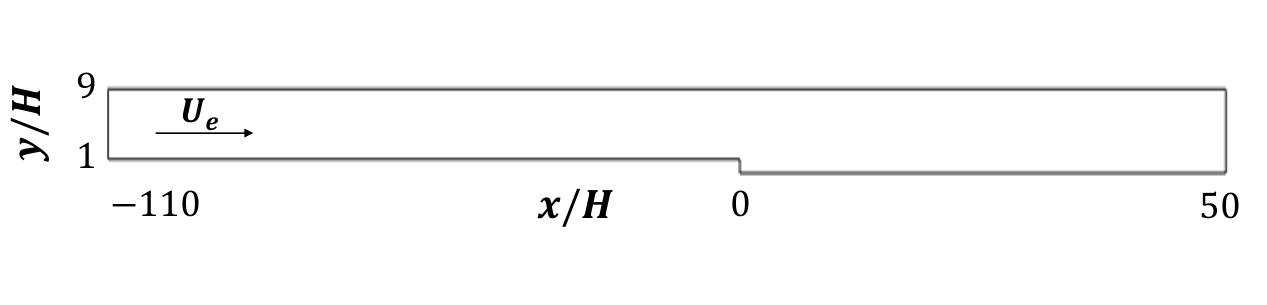}
    \caption{A schematic of the flow over a backward-facing step studied by \citet{driver1985features}. The flow develops over a flat plate upstream of the step before separating at the corner. The flow eventually reattaches downstream with its reattachment length shown to be a function of the Reynolds number in \citet{jovic1994backward}. }
    \label{fig:bfsgeom}
\end{figure}

\noindent
The experimental configuration of $E_r \approx 1.1$ and $Re_H = U_{\infty} H/\nu = 37500 $ (where $H$ is the step height) from \citet{driver1985features} is simulated in this work. Fig. \ref{fig:bfsgeom} provides a schematic of the exact simulation setup with the inlet extended significantly upstream (at $x/H=-110$) to develop a realistic turbulent boundary layer that matches the experimental $Re_{\theta}$ at the upstream reference location of $x/H=-4$. The inviscid, top boundary is placed at $y/H=9$ to match the expansion ratio; an equilibrium wall model closure is applied on the bottom wall. A separate set of experiments in \citet{jovic1994backward} previously suggested only limited differences in the pressure distribution associated with a double-sided expansion and a single-sided expansion at the same expansion ratio. The outlet is placed at $x/H=50$ with non-reflective boundary conditions. Table \ref{table:bfstable} provides the details of the gridding approach for this flow. A schematic of the employed computational mesh is available in Appendix III.

\begin{table}
\centering
\begin{tabular}{ p{2cm}p{1.5cm}p{2cm}p{2cm}p{3cm}p{2cm} }
Mesh & $N_{cv}$ &  $max. \Delta / H$  & $min. \Delta / H$ & $x_r/H$ \\
\hline\noalign{\vspace{3pt}}
Coarse  & $2$ Mil.  &  $1$ & $1.2 \times 10^{-1}$ & 5.04 \\
Medium  & $8$ Mil.    &  $1$ & $6.2 \times 10^{-2}$ & 5.56 \\
Fine    & $32$ Mil.   & $1$ & $3.1 \times 10^{-2}$ & 6.24 \\
% Very Fine    & $130$ Mil.   & $1$ & $1.6 \times 10^{-2}$ & 6.24 \\
\end{tabular}
\caption{Mesh parameters for the $Re_{H} = 37500$ case of the flow over the backward-facing step studied by \citet{driver1985features}. The last column denotes the mean flow reattachment length as a function of the employed grid. The reported experimental value is $x_r/H = 6.26 \pm 0.05$. }
\label{table:bfstable}
\end{table}

\begin{figure}[!ht]
    \centering
    \includegraphics[width=1\columnwidth]{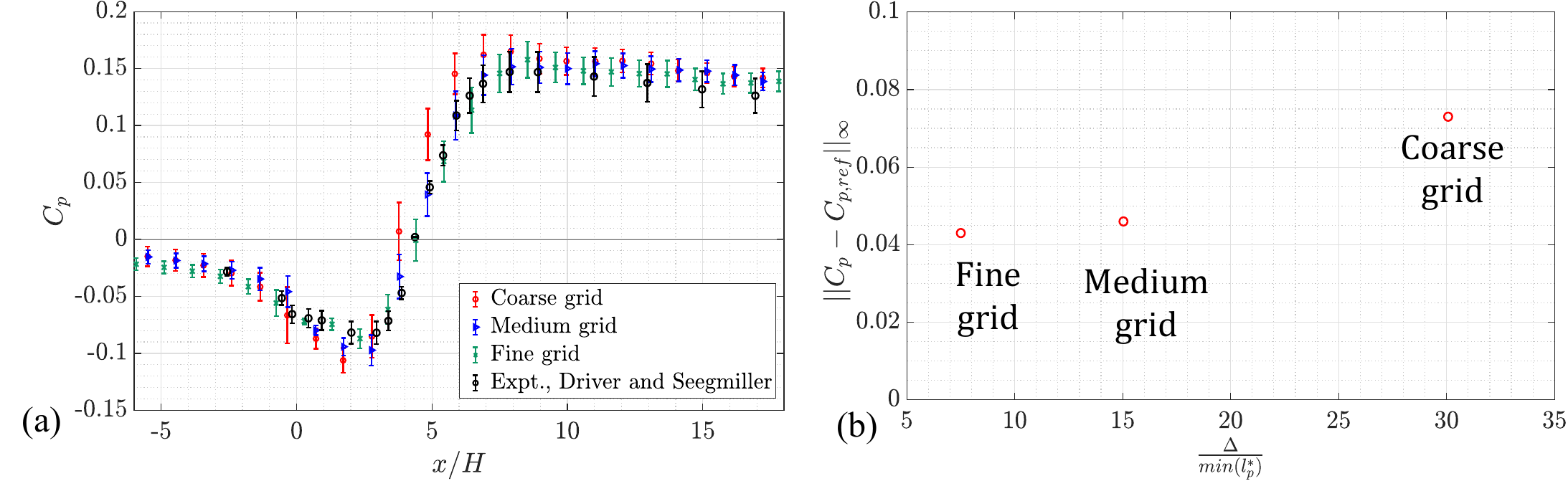}
    \caption{The (a) bottom wall surface pressure for the flow past a backward-facing step studied by \citet{driver1985features} at a step-height Reynolds number, $Re_{H}=37500$. Subfigure (b) provides the error convergence plot upon grid-refinement at this Reynolds number. In the experiments, the flow separates past the step and reattaches with the bottom wall at $x_r/H \approx 6.26$. The scaling variable $l^*_p$ is the pressure gradient length scale imposed by the flow upstream of the separation point ($x/H=0$) for consistency with the other cases considered in this work. The vertical bars in the simulations and the experiment represent the uncertainty around the mean. In the simulations, the root mean square error is plotted for the same.  }
    \label{fig:bfscp37500}
\end{figure}

\begin{figure}[!ht]
    \centering
    \includegraphics[width=1\columnwidth]{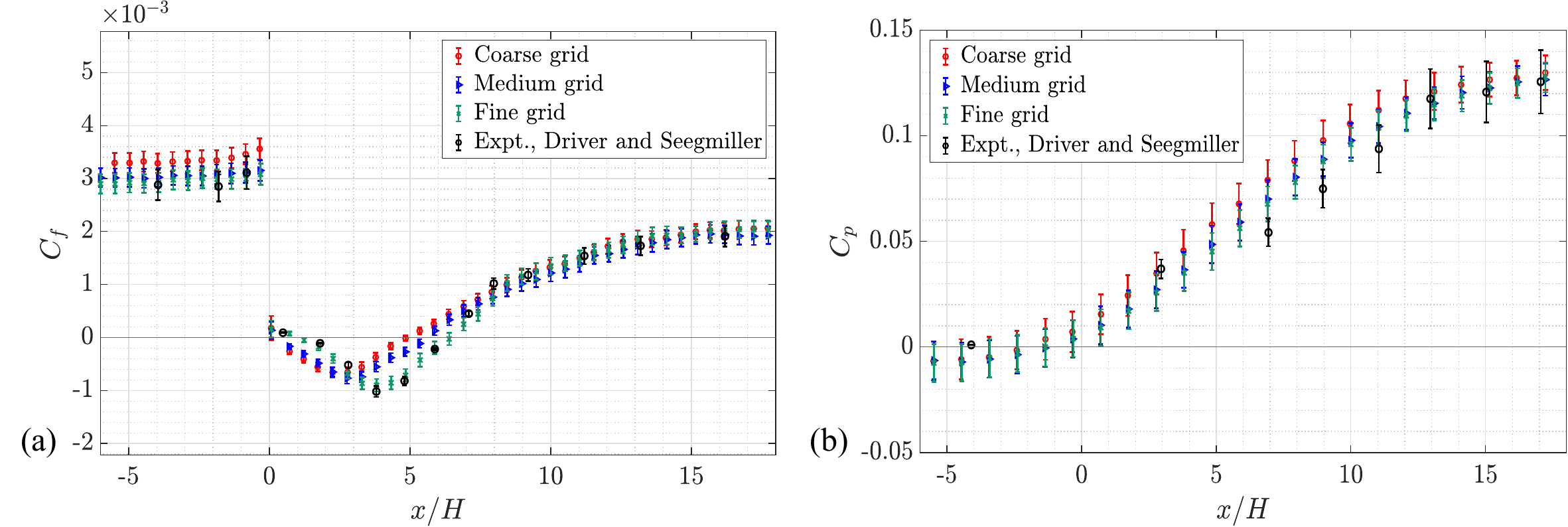}
    \caption{The (a) skin-friction distribution and the (b) surface pressure distribution at the top wall for the flow past a backward-facing step studied by \citet{driver1985features} at a step-height Reynolds number, $Re_{H}=37500$. The vertical bars in the simulations and the experiment represent the uncertainty around the mean. In the simulations, the root mean square error was used to quantify this uncertainty. ` }
    \label{fig:bfscp37500_2}
\end{figure}

\noindent
Fig. \ref{fig:bfscp37500}(a) compares the predictions of the surface pressure on the bottom wall with the experiments \citep{driver1985features}. The predictions of the surface pressure improve on grid refinement with good agreement with the experiment resulting when using the finest grid. In figure \ref{fig:bfscp37500}(b), the error in $C_p$ converges to $\approx 0.04-0.05$ when $\Delta/min(l_p^*) \sim 7 - 15$. In Fig. \ref{fig:bfscp37500_2}(a), the skin-friction distribution is compared against the experiments, and the finest grid recovers the experimental reporting well in the separated, and recovery regions. In Fig. \ref{fig:bfscp37500_2}(b), the pressure at the top wall is also captured well in wall-modeled LES on the finest grid. Overall, for this flow, our results suggest that the grid-resolution requirement for the accurate prediction of the reattachment point ($x_r$), and the surface pressure ($C_p$)  past a step are similar to the resolution required to accurately capture the extent of the separated flow region over smooth surfaces. 

% The grid point scaling with Reynolds number is analyzed by performing wall-modeled LES at a reduced $Re_H = 6800$ as well. Fig. \ref{fig:bfscp6800} shows the results of $C_p$ and $C_f$ on the bottom wall at this Reynolds number. In the absence of experimental results, wall-modeled LES predictions on a ``very fine" grid are used as a reference. The fine-grid results agree with the very fine grid, implying grid-convergence of LES, and that the Reynolds number scaling of required grid points for wall-modeled LES of this flow scale with $min(l_p)$ as well. 

% \begin{figure}[!ht]
%     \centering
%     \includegraphics[width=1\columnwidth]{bfsReh6800.pdf}
%     \caption{The (a) surface pressure at the bottom wall, and the (b) skin-friction distribution plots for the flow past a backward-facing step studied at a reduced step-height based Reynolds number, $Re_{H}=6800$. The bars on the simulation results denote the root mean square variations in the predictions about the mean value of $C_p$ and $C_f$.  }
%     \label{fig:bfscp6800}
% \end{figure}

\section{Concluding Remarks}
In this work, the Reynolds number scaling of the required number of grid points to perform wall-modeled LES of flow encountering separation is examined. Based on the various time scales in a turbulent boundary layer experiencing external pressure gradients, it is suggested that the Reynolds number dependence of the ``under-equilibrium" length scale (scales for which the viscous and the pressure gradient effects approximately balance each other) can be hypothesized to be $l_p \sim Re^{-2/3} $. The same scaling is revealed based on a simplified Green's function type solution of the pressure equation around the separation point. \emph{A-priori} analysis suggests that the minimum resolution required to reasonably predict the shear stress (errors lower than approximately $10-15\%$ in the entire domain) with the equilibrium wall model is at least $ \mathcal{O}(10) $$l_p$ irrespective of the Reynolds number and the Clauser parameter for several non-equilibrium flat-plate boundary layers and airfoil flows. Several \emph{a-posteriori} calculations are then performed to determine the accuracy of this scaling. Numerical simulations are performed and compared with experimental data for the flow over the Boeing speed bump, a diffuser, Notre-Dame Ramp, and the backward-facing step. The results suggest that for these flows, scaling the grids ($\Delta$) to maintain a fixed $\Delta/l_p$ results in accurate predictions of flow separation at the same ``nominal" grid resolution across different Reynolds numbers. Upon employing a dynamic subgrid-scale model and an equilibrium wall modeling closure, the minimum resolution requirement for capturing the flow separation was found to be $\Delta/min(l_p) \sim \mathcal{O}(10)$. We also find that selected flow cases that use existing non-equilibrium wall models also require similar grid resolutions to achieve accurate results (as detailed in Appendix I). Finally, it is suggested that in more complex, three-dimensional flows, at least, locally, the grid-point requirements to predict flow-separation may scale as $Re^{4/3}$, which is more restrictive than the previously proposed flat-plate boundary layer-based estimates for wall-modeled LES.

\section*{Appendix I: Performance of existing non-equilibrium wall models}
%\vspace{-5pt}
Several non-equilibrium wall models have been proposed \citep{kawai2013dynamic,park2017wall,zhou2023large,kamogawa2023ordinary} to account for the flow non-equilibrium by modeling the pressure gradient and convective terms either using governing equations, or data-driven methods. These studies have all reported some improvements in the \emph{a-posteriori} wall-modeled LES predictions upon employing their respective non-equilibrium wall models over a range of mildly separated flows. Table \ref{table:noneqbmodels} summarizes the near-wall, wall-normal grid resolution (non-dimensionalized by the viscous scale imposed by the strongest pressure gradient) required when using these models to predict the quantities of interest accurately. It is clear that for all considered flows, the required resolution to predict the quantities of interest accurately is comparable, and not significantly coarser, than the resolution required with the equilibrium wall model as found in this work [$\Delta/min(l_p) \sim \mathcal{O}(10)$]. Further research may provide the Reynolds number scaling of these models, and improvements in the $N_{cv} \sim Re^{4/3}$ cost scaling, if any.   

\begin{table}
\centering
\begin{tabular}{ p{6.5cm}p{4cm}p{1.6cm} }
Model Description & Flow & $ \Delta/min. (l_p)$  \\
\hline\noalign{\vspace{3pt}}
PDE, non-equilibrium & NASA Hump & $\approx 16$ \\ 
wall-model \citep{park2017wall} &   &  \\
PDE, non-equilibrium & Airfoil & $\approx 14$ \\ 
wall-model \citep{kawai2013wall} & & \\
Reinforcement learning & Periodic Hill & $\approx 14$ \\ wall-model \citep{zhou2023large} && \\
ODE, non-equilibrium & Suction, blowing  & $\approx 15$ \\ 
wall-model \citep{kamogawa2023ordinary} & boundary layer&\\
\end{tabular}
\caption{The minimum wall normal grid resolutions reported for accurate prediction of quantities of interest upon leveraging existing non-equilibrium wall models for a range of mildly separated flows.  }
\label{table:noneqbmodels}
\end{table}

\section*{\label{sec: Grid point estimates} Appendix II: Grid estimates of wall-modeled LES based on fixed outer resolution ($\frac{\Delta}{\delta}$)}
%\vspace{-5pt}

\noindent
Previous efforts \citep{chapman1979computational,choi2012grid,yang2021grid} have provided grid-point estimates for wall-modeled LES assuming that the grid resolutions scale in outer flow units ($\delta$). It is known that the mean local shear determines the dynamics of turbulence at different wall-normal locations in both equilibrium and non-equilibrium flows \citep{flores2007vorticity,lozano2019characteristic}. As long as the pressure gradient is not too strong, this shear rate also does not significantly vary across the outer region of the boundary layer for non-equilibrium flows. Hence, a similar assumption about requiring a fixed outer resolution ($\delta/\Delta$) of the boundary layer may be motivated for non-equilibrium flows. It is highlighted, however, that such a scaling relies on the assumption that the subgrid-scale and wall models remain accurate in a range of non-equilibrium conditions. 

\noindent
For this purpose, we recall the relationships suggested in \citet{agrawal2023thwaites} between the growth rates of the momentum thickness ($\theta$) and the boundary layer thickness ($\delta_{99}$). For a turbulent boundary layer with freestream velocity distribution, $U_e (x)$, the authors proposed that 
\begin{equation}
 \frac{d \delta }{d x} \approx \frac{5}{200} + \frac{8}{200} m - \frac{U_e}{200 \nu } \frac{d \theta^2}{d x}   +  \frac{d \delta{zpg,corr}} {dx}.
\label{eqn:ddeltadx}
\end{equation}
where $\delta_{zpg,corr}/x = 0.16 Re^{-1/7}_x = 0.16 (U_e x / \nu )^{-1/7} $. $m$ is the Holstein-Bohlen parameter given as, $m=\theta^2/\nu dU_e/dx$. The expression for the growth of the momentum thickness is another ordinary differential equation given as 

\begin{equation}
    \frac{d}{d x} (U^{C_m}_e \theta^2) \approx  \nu C_c U^{C_m-1}_e   +  C_{Re, \infty} {U^{C_m}_e \theta }.  
    \label{eqn:mymodel}
\end{equation}
where $C_c = 1.45$, $C_m = 7.2$, and $C_{Re, \infty} = 0.0024$. 
Consider, a turbulent boundary layer experiencing mild pressure gradients, such that it grows according to Equations \ref{eqn:ddeltadx}-\ref{eqn:mymodel}.  Let $\eta = \frac{x^*}{U^2_e}\frac{dP}{\rho dx}$ be the parameter that governs the streamwise distance over which the pressure gradient grows slowly (or remains locally constant). Bernoulli's principle implies that over the region ($0 - x^{*}$) 
\begin{equation}
    U_e (x) = U^0_e \sqrt{1 - 2  \frac{x^*}{U^{2,0}_e}\frac{dP}{\rho  dx} } \approx U^0_e ( 1 -  \eta ) .
\end{equation}
Using this relation and the equations above, it can be shown that to the leading order (in $\eta$) 
 \begin{equation}
 \frac{d \theta }{d x}  \approx  \left[\frac{C_c \nu }{2 U^0_{e} \theta } + \frac{C_{Re, \infty}}{2}\right] +  
  \left[\eta \left( \frac{C_c \nu }{2  U^0_e  \theta} + \frac{C_m \theta}{2 x} \right)  \right].
 \label{eqn:dthetadxsplit}
 \end{equation}

\noindent
Due to the proposed linear expansion in ($m, Re_{\theta}$) space for the growth of the momentum thickness, the effect of the Reynolds number and the pressure gradient on the growth of the boundary layer are explicitly separable. The terms in the first parenthesis represent the growth rate of a zero-pressure gradient boundary layer. Thus 

 \begin{equation}
 \frac{d \theta }{d x} \approx   \frac{d \theta^{zpg} }{d x} 
 + \left[ \eta \left( \frac{C_c \nu }{2  U^0_e  \theta} + \frac{C_m \theta}{2 x} \right)  \right].
 \label{eqn:dthetadxsplit2}
 \end{equation}

\noindent
If the inviscid pressure gradient distribution, $\eta = \eta(x)$, was known, then Equation \ref{eqn:dthetadxsplit2} could be integrated to give estimates for the growth of the boundary layer.  If we further assume that the pressure gradients are sufficiently mild such that, $\theta \approx \theta^{zpg} + \eta \frac{d\theta}{d\eta}\big|_{zpg}$, then the leading order expansion in $\eta$ would yield 
 
 \begin{equation}
 \frac{d \theta }{d s} \approx (0.016 + 0.6 \eta) Re^{-1/7}_x.
 \label{eqn:dthetadxsplit3}
 \end{equation}

\noindent
Using Equations \ref{eqn:dthetadxsplit3} and \ref{eqn:ddeltadx}, an approximate relationship between the boundary layer thickness, $\delta$ (evaluated using the method of \citet{griffin2021general}) in ($\eta, Re_x$) space can be derived as 
\begin{equation}
    \delta (x) \approx \delta^{zpg} + 10^{-5} \eta x   Re^{13/14}_x \approx 0.16 x Re^{-6/7}_x  +  10^{-5} x [\frac{x}{ U^{2,0}_e }  \frac{  dP}{\rho dx}] Re^{13/14}_x .
    \label{eqn:deltafinalwmles}
\end{equation}

\noindent
In the asymptotic limit of a large Reynolds number, the model predicts that the boundary layer growth is determined linearly by the pressure-gradient, and slightly sub-linearly by the Reynolds number. Following \citet{choi2012grid,yang2021grid}, an integral relationship between $\delta$ and the total number of points required for performing wall-modeled LES is written as  
 \begin{equation}
     N_{pts}^{wm} = N(x<x_0) + \int^{L_x}_{x_0}\int^{L_z}_{0} \frac{n_x n_y n_z }{\delta^2} dx dz .
     \label{eqn:choimoin}
 \end{equation}
 where $n_x, \; \mathrm{and} \; n_z$ are the number of points inside the boundary layer in the streamwise, and spanwise directions respectively. $x_0, \; \mathrm{and} \; L_x,$ are the limits of the streamwise integral, and $L_z$ is the domain size in the spanwise direction. Thus, for a given inviscid flow distribution, the grid-point requirements for a boundary layer experiencing mild pressure gradients can be evaluated using Equations \ref{eqn:deltafinalwmles} and \ref{eqn:choimoin}. 

\section*{Appendix III: Details of computational meshes}
The computational grids employed in this work are based on a centroidal Voronoi diagram generated from a hexagonally close-packed lattice of seed points. While refining the grid, the control volumes are refined isotropically by factors of two in the layers near the boundaries. Each refinement step injects a thin layer (containing 10 cells) into the regions immediately adjacent to the wall of twice as fine cells in each direction compared to the previous mesh. Further, Lloyd iterations are performed to make the mesh transitions between layers of different cell sizes smoother. Control volumes far from domain boundaries and refinement transitions consist of tessellated truncated octahedra. Although not required, the mesh arrangement was chosen to be uniform in the streamwise and spanwise directions to properly evaluate the predictive abilities of wall-modeled LES. 

\begin{figure}[!ht]
    \centering
    \includegraphics[width=0.7\columnwidth]{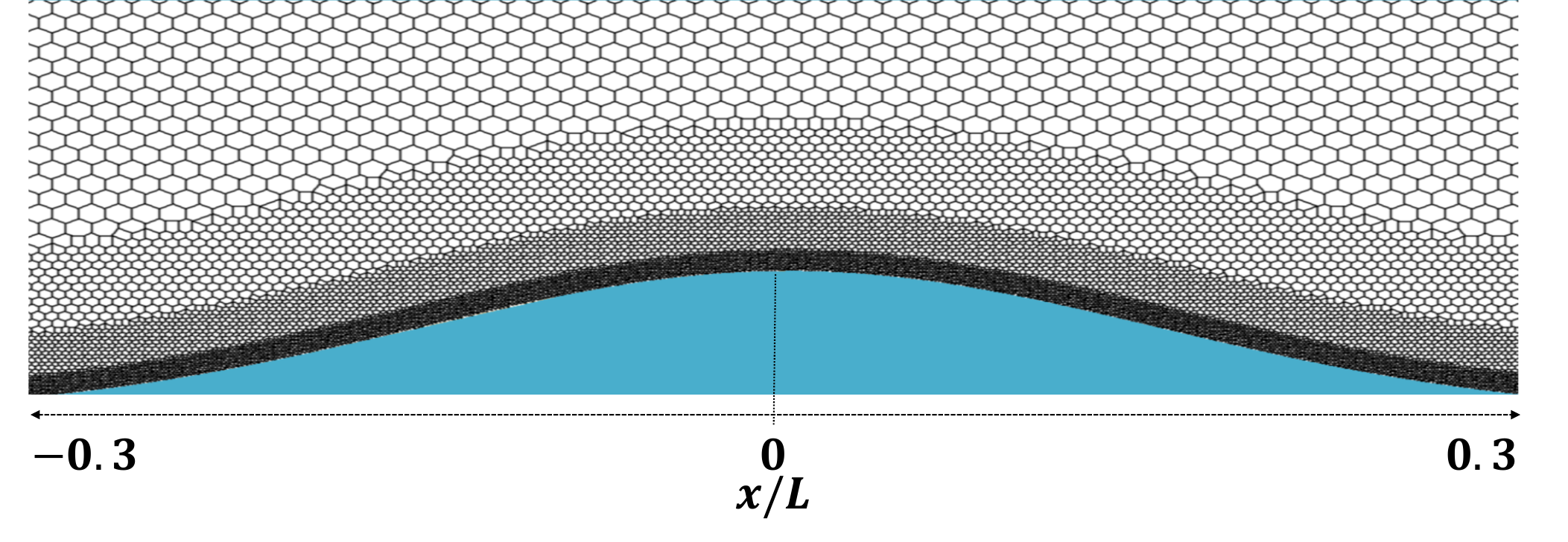}
    \caption{Cross section of the ``coarse" grid for the spanwise-periodic variant geometry of the Boeing speed bump \citep{uzun2021high} at $Re_L = 2 \times 10^6$. The streamwise extent is $x/L \in [-0.3,0.3]$ showing both the fore and the aft sides of the bump. Only the lower half of the vertical extent of the domain is shown. The top half is meshed uniformly up to the top wall. Three layers of isotropic refinement are visible adjacent to the bump surface. The qualitative arrangement of the control volume remains identical to this arrangement while simulating the higher Reynolds number flows at $Re_L = 5, \; 10 \times 10^6$.  }
    \label{fig:boeingmesh}
\end{figure}

\begin{figure}[!ht]
    \centering
    \includegraphics[width=0.7\columnwidth]{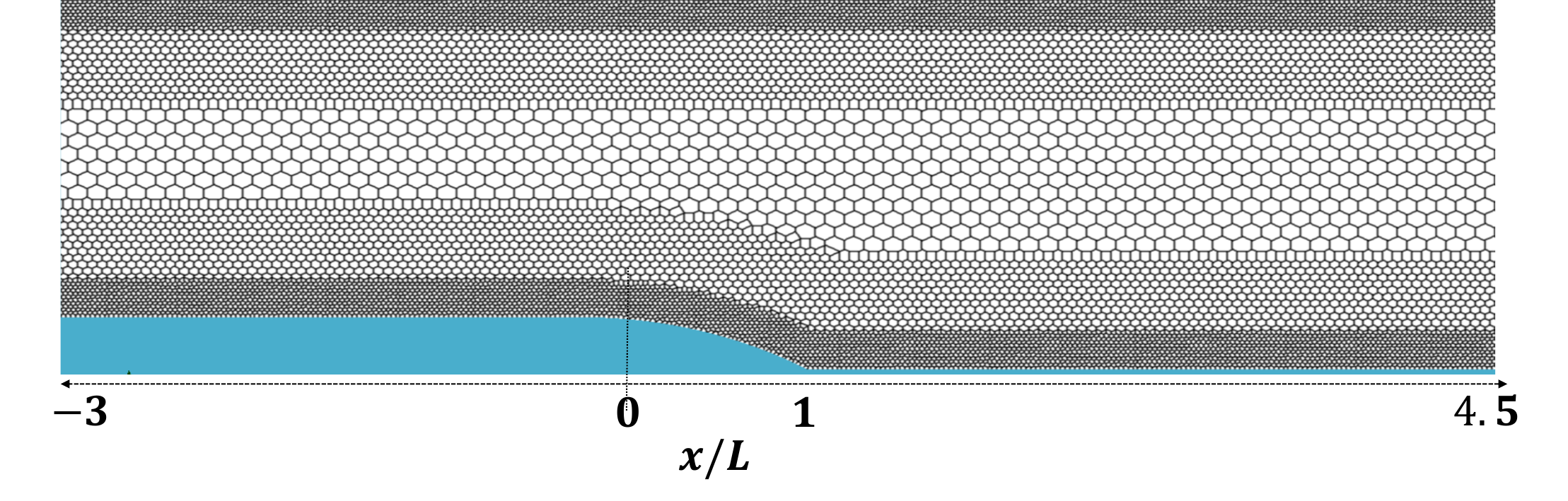}
    \caption{Cross section of the ``coarse" grid for the Song-Eaton diffuser geometry  \citep{song2004reynolds} at $Re^{ref}_{\theta} = 3400$. The streamwise extent is $x/L \in [-3,4.5]$ showing both the upstream (of the diffuser) and the downstream regions. The entire vertical extent of the domain is shown. Two layers of isotropic refinement are visible adjacent to the walls on both the lower and the upper walls. The qualitative arrangement of the control volume remains identical to this arrangement while simulating the higher Reynolds number flows at $Re^{ref}_{\theta}  = 13200, \; 20100$.  }
    \label{fig:songmesh}
\end{figure}

\begin{figure}[!ht]
    \centering
    \includegraphics[width=0.7\columnwidth]{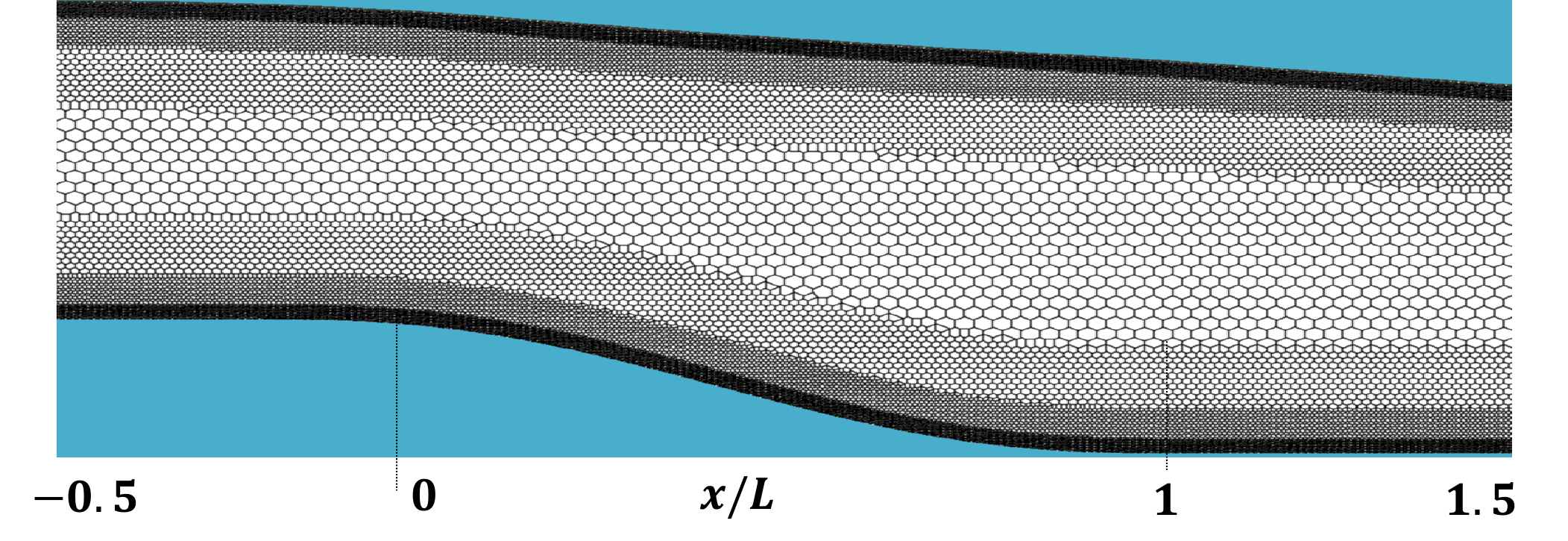}
    \caption{Cross section of the ``coarse" grid for the smooth-ramp geometry  \citep{simmons2020experimental} at $Re_{H} = 8.4 \times 10^5$. The streamwise extent is $x/L \in [-0.5,1.5]$ showing both the upstream (of the ramp) and the downstream flat-plate regions; the entire vertical extent of the domain corresponding to the streamwise region, $x/L \in [-0.5,1.5]$, is also shown. Three layers of isotropic refinement are visible adjacent to the walls on both the lower and the upper walls. }
    \label{fig:simmonsmesh}
\end{figure}

\begin{figure}[!ht]
    \centering
    \includegraphics[width=0.78\columnwidth]{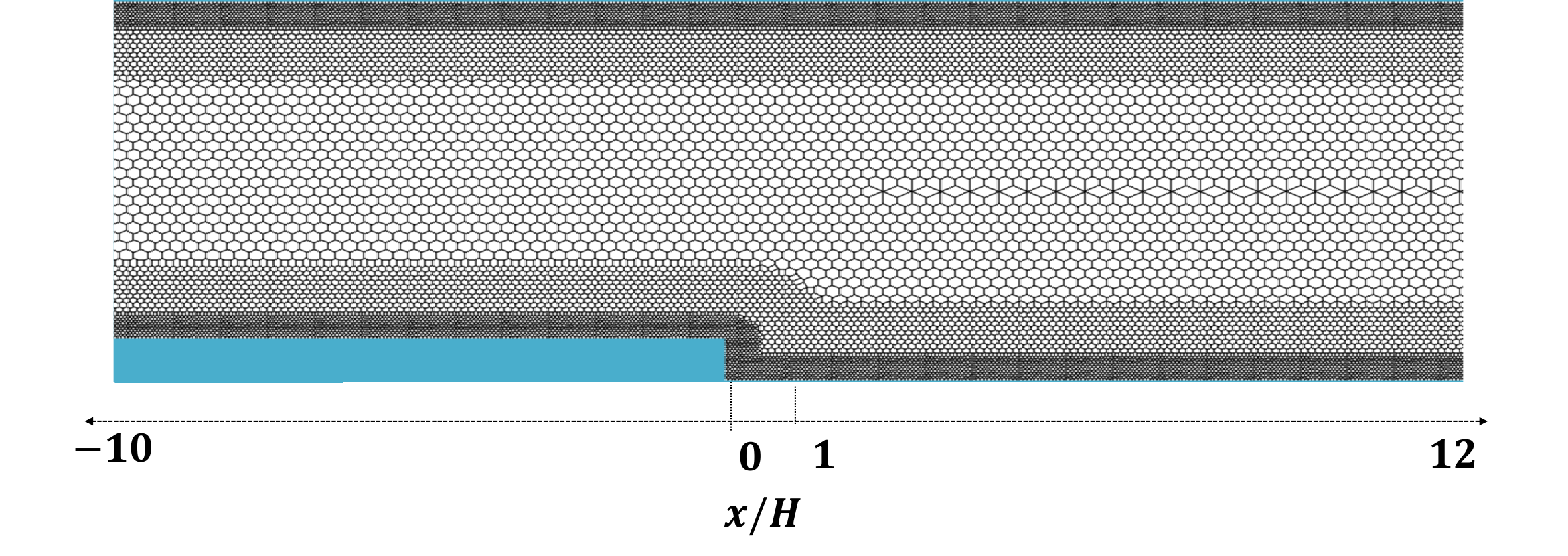}
    \caption{Cross section of the ``coarse" grid for the backward-facing step geometry  \citep{jovic1994backward} at $Re_{H} = 37500$. The streamwise extent is $x/L \in [-10,12]$ and the entire vertical extent of the domain is shown. Three layers of isotropic refinement are visible adjacent to the walls on both the lower and the upper walls.}
    \label{fig:bfsmesh}
\end{figure}

The schematics of the meshes corresponding to the ``coarse grid" for each flow case are now provided as follows. Fig. \ref{fig:boeingmesh} corresponds to the mesh distribution for the flow over the Boeing speed bump. Similarly, Figures \ref{fig:songmesh}, \ref{fig:simmonsmesh}, and \ref{fig:bfsmesh} correspond to the Song-Eaton diffuser, Notre-Dame smooth ramp, and the backward-facing step cases respectively. 


\section*{Appendix IV: Effect of subgrid-scale model}
%\vspace{-5pt}

\begin{figure}[!ht]
    \centering
    \includegraphics[width=0.9\columnwidth]{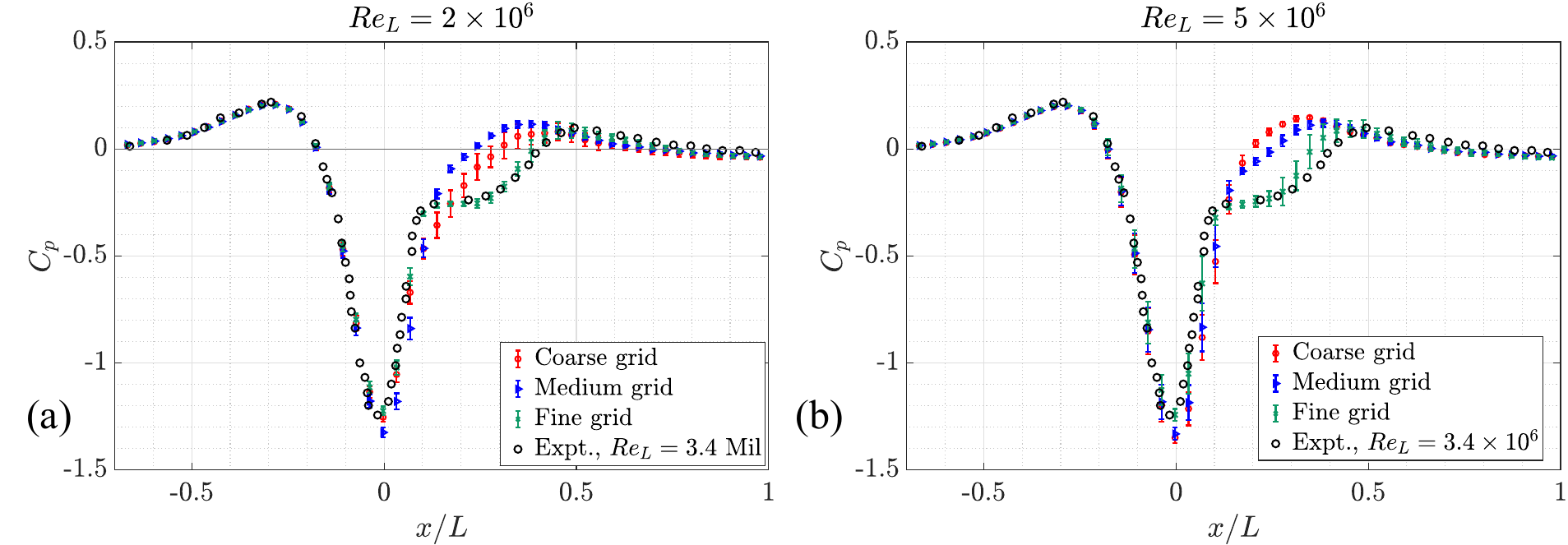}
    \caption{The streamwise distribution of the surface pressure coefficient, $C_p$ for (a) $Re_L = 2 \times 10^6$ and (b) $Re_L = 5 \times 10^6$ for the flow over the Boeing speed bump using the dynamic Smagorinksy model and the equilibrium wall model. Note that across the two Reynolds numbers, the nominal ``coarse", ``medium" and ``fine" grids are scaled in their resolution by a factor of $Re_L^{2/3}$. The vertical bars in the simulations represent the root mean square error around the mean value of $C_p$.   }
    \label{fig:speedbumpdsm}
\end{figure}

Fig. \ref{fig:speedbumpdsm} compares the predictions of $C_p$ upon leveraging the dynamic Smagorinsky model \citep{moin1991dynamic} instead of the dynamic tensor-coefficient model \citep{agrawal2022non}. As reported previously in \citet{agrawal2022non}, on the lower Reynolds number, $Re_L = 2 \times 10^6$, the predictions of wall-modeled LES using the dynamic Smagorinsky model converge towards the reference experiments non-monotonically. However, at the $Re_L \uparrow$ up to $Re_L = 5 \times 10^6$, the predictions approach the solution monotonically. For both Reynolds numbers, the ``fine" grid is required for wall-modeled LES to match the experimental measurements, implying that the effect of the subgrid-scale model is minimal in improving the Reynolds number scaling of grid points required in wall-modeled LES of this flow. 

\section*{Acknowledgments}
%\vspace{-5pt}
This work was supported by NASA's Transformational Tools and Technologies project under grant number \\ \#80NSSC20M0201. Computing resources were awarded through the Oak Ridge Leadership Computing Facility (DoE ALCC). We thank Michael P. Whitmore for the helpful comments on the initial version of this manuscript. We also acknowledge fruitful discussions with Ahmed Elnahhas, Kevin Griffin, and Christoffer Hansen.

\bibliographystyle{unsrt}

\end{document}